\documentclass[10pt,journal,compsoc]{IEEEtran}
\ifCLASSOPTIONcompsoc
  \usepackage[nocompress]{cite}
\else
  \usepackage{cite}
\fi
\ifCLASSINFOpdf
\else
\fi
\usepackage{amsmath,amssymb,amsfonts,url}
\usepackage{algorithmic}

\usepackage{graphicx}
\usepackage{textcomp}
\usepackage{xcolor}

\usepackage[linesnumbered]{algorithm2e}
\usepackage{tabularx,ragged2e,booktabs,caption}
\usepackage{balance}
\usepackage{multirow}

\ifCLASSOPTIONcompsoc
\usepackage[caption=false,font=normalsize,labelfont=sf,textfont=sf]{subfig}
\else
\usepackage[caption=false,font=footnotesize]{subfig}
\fi

\SetAlCapSkip{2ex}

\DeclareMathAlphabet{\mathsl}{OT1}{cmr}{m}{sl}
\DeclareMathAlphabet{\mathsc}{OT1}{cmr}{m}{sc}
\DeclareMathAlphabet{\mathslbf}{OT1}{cmr}{bx}{sl}

\begin{document}
%
\title{Performance Evaluation and Modeling of Cryptographic Libraries
    for MPI Communications}
%
%
%
%

\author{Abu~Naser,~\IEEEmembership{}
        {Mehran~Sadeghi~Lahijani},~\IEEEmembership{} 
	Cong~Wu,~\IEEEmembership{} 
	Mohsen~Gavahi,~\IEEEmembership{} \\
	Viet~Tung~Hoang,~\IEEEmembership{}
	Zhi~Wang,~\IEEEmembership{}
        and~Xin~Yuan~\IEEEmembership{}

\IEEEcompsocitemizethanks{\IEEEcompsocthanksitem A. Naser, M. S. Lahijani, C. Wu, M. Gavahi, V. T. Hoang, Z. Wang, and X. Yuan are with
    Department of Computer Science, Florida State University, Tallahassee, FL 32306-4530.\protect\\
    E-mail: \{naser, sadeghil,  wu, gavahi, tvhoang, zwang, xyuan\}@cs.fsu.edu
}
\thanks{}}        

\IEEEtitleabstractindextext{%
\begin{abstract}

  In order for High Performance Computing (HPC) applications
  with data security requirements to execute in the public cloud,
  the cloud infrastructure must ensure privacy and integrity of data. 
%
To meet this goal, we consider incorporating
  encryption 
	in the
  Message Passing Interface (MPI) library. 
	We empirically evaluate four contemporary cryptographic
  libraries, OpenSSL, BoringSSL, Libsodium, and CryptoPP using
  micro-benchmarks and NAS parallel benchmarks on two different
  networking technologies, 10Gbps Ethernet and 40Gbps InfiniBand. 
  We 
	also develop accurate models that 
  allow us to reason about
  the performance of encrypted MPI communication in different situations, 
  and give guidance on how to improve encrypted MPI performance.

\end{abstract}

}

\maketitle

\IEEEdisplaynontitleabstractindextext

%
\IEEEpeerreviewmaketitle

\IEEEraisesectionheading{\section{Introduction}\label{sec:introduction}}

\IEEEPARstart
There is a current trend that more and more High Performance Computing
(HPC) applications are moved to execute in the public cloud. Some of these
applications process sensitive data, such as medical, financial,
and engineering documents. In order for such HPC applications to
run on the cloud, the cloud infrastructure for HPC must provide
privacy and integrity. A very large number of
HPC applications are developed based on
the Message Passing Interface (MPI) library, the de facto
communication library for message passing applications.
Incorporating privacy and integrity mechanisms in the MPI library
will allow MPI applications to run in the public cloud with strong
security guarantees.

The existing efforts to secure MPI libraries with
encryption contain severe security flaws. For example,
ES-MPICH2~\cite{Ruan:2012:EMP:2197079.2197242}, the first such MPI
library, uses the weak ECB 
(Electronic Codebook) mode of operation that has known
vulnerabilities~\cite[page 89]{Book:KL14}. 
In addition, no existing
encrypted MPI libraries provide meaningful data integrity, 
meaning that 
data could potentially be modified without being detected. Hence,
it is urgent to 
apply modern cryptographic theory and practice to
properly secure MPI communications. 

In recent years, significant efforts have been put in the hardware
and software to improve the performance of security operations.
Recent processors from all major CPU vendors have introduced
hardware support for cryptographic operations (e.g., Intel AES-NI
instructions to accelerate the AES algorithm, 
or the x86 CLMUL instruction set to improve the speed of
finite-field multiplications). Modern cryptographic libraries,
including OpenSSL~\cite{openssl}, BoringSSL~\cite{boringssl}, 
Libsodium~\cite{libsodium} and CryptoPP~\cite{cryptopp}, all support
hardware-accelerated cryptographic operations. All of these libraries
received intensive security review; and some (OpenSSL and CryptoPP)
even passed the Federal Information Processing Standards
(FIPS) 140-2 validation. These 
libraries have different usability, functionality, and performance.
Moreover, recent technology advances have shifted the communication
bottleneck from the network links to the network end-points.
As a result, the computationally intensive encryption operations
may introduce significant overheads to MPI communications
when security mechanisms are incorporated in the MPI library. It is
thus critical find out which library is the best option
for MPI communications, and to understand the overheads introduced
due to the cryptographic operations. 

In this work, we develop 
encrypted MPI libraries that are built
on top of four cryptographic libraries OpenSSL, BoringSSL, Libsodium,
and CryptoPP. Using these libraries, we empirically evaluate the
performance of encrypted MPI communications with micro benchmarks and NAS 
parallel benchmarks \cite{Bailey:1991:NPB:2748645.2748648} 
on two networking technologies, 10Gbps Ethernet and 40Gbps InfiniBand QDR.
The main conclusions include the following:

\begin{itemize}

\item Different cryptographic libraries result in very different
  overheads. Specifically, OpenSSL and BoringSSL are on par with each other;
  and their performance is much higher than that of Libsodium and CryptoPP,  on both  Ethernet
  and InfiniBand.

\item For individual communication, encrypted MPI introduces relative small overhead
  for small messages and large overhead for large messages. For example, 
  on the 10Gbps Ethernet, even for BoringSSL, 
	encrypted MPI under the AES-GCM encryption scheme~\cite{Dworkin:2001:SER:2206247} 
	introduces 5.9\% overhead for 256-byte messages and 
  78.3\% for 2MB messages in the ping-pong test. On the 40Gbps InfiniBand,
  BoringSSL introduces 80.9\% overhead
  for 256-byte messages and 215.2\% overhead for 2MB messages
  in the ping-pong test. 
  This calls for developing new techniques to optimize the combination of
  encryption and MPI communications. 
\item For more practical scenarios, the cryptographic overhead is not as significant. 
    On average, BoringSSL only introduces 
    12.75\% overhead on Ethernet and 17.93\% overhead on InfiniBand
  for NAS parallel benchmarks (class C running on 64 processes and 8 nodes).
\end{itemize}

To gain a theoretical understanding of the performance of encrypted MPI
point-to-point communication, we investigate the modeling of
encrypted MPI point-to-point communication in two settings: a single
point-to-point communication like the Ping-Pong test and
multiple concurrent point-to-point communications like the OSU Multiple-Pair
benchmark \cite{OSUBM}. We show that the performance of encrypted MPI
point-to-point communication can be accurately modeled using the
Hockney model~\cite{Hockney94} that has been used
to model MPI
communication~\cite{Thakur05,Faraj05,Rabenseifner04,Faraj2008,PATARASUK2008809,Patarasuk:2009}, and the max-rate model of Gropp et
al.~\cite{gropp2016modeling} that was proposed to model concurrent
communications. Moreover, the parameters of our models can be derived from
independent benchmarking of conventional MPI and encryption libraries. 
The simplicity of our models offers a simple and intuitive explanation
for the performance of encrypted MPI point-to-point communication
across different cryptographic libraries and networks, and allows for
accurate prediction of encrypted MPI point-to-point communication
performance as technology advances. Using these models, we reason about
the performance of encrypted MPI point-to-point communication in 
different situations and discuss the potential benefits of some optimization 
techniques to improve the performance of encrypted MPI communication.

\newcommand{\Keys}{\mathrm{Keys}}
\newcommand{\Dom}{\mathrm{Dom}}

\newcommand{\secref}[1]{Section~\ref{#1}}
\newcommand{\apref}[1]{Appendix~\ref{#1}}
\newcommand{\thref}[1]{Theorem~\ref{#1}}
\newcommand{\sthref}[1]{Th.~\ref{#1}}
\newcommand{\defref}[1]{Definition~\ref{#1}}
\newcommand{\scorref}[1]{Cor.~\ref{#1}}
\newcommand{\lemref}[1]{Lemma~\ref{#1}}
\newcommand{\clref}[1]{Claim~\ref{#1}}
\newcommand{\propref}[1]{Proposition~\ref{#1}}
\newcommand{\factref}[1]{Fact~\ref{#1}}
\newcommand{\remref}[1]{Remark~\ref{#1}}
\newcommand{\obsref}[1]{Observation~\ref{#1}}
 \newcommand{\figref}[1]{\textmd{Fig.}~\ref{#1}}
\newcommand{\tabref}[1]{Table~\ref{#1}}
\newcommand{\consfigref}[1]{Construction~\ref{#1}}
\renewcommand{\eqref}[1]{\mbox{Eq.~(\ref{#1})}}
\newcommand{\IV}{\mathrm{IV}}
\newcommand{\concat}{{\,\|\,}}

\newcommand{\getsr}{{\:{\leftarrow{\hspace*{-3pt}\raisebox{.75pt}{$\scriptscriptstyle\$$}}}\:}}
\newcommand{\Procedure}{\textbf{procedure }}
\newcommand{\Enc}{\texttt{Enc}}
\newcommand{\Dec}{\texttt{Dec}}
\newcommand{\Gen}{\texttt{Gen}}
\newcommand{\sendbuf}{\mathsl{sendbuf}}
\newcommand{\recvbuf}{\mathsl{recvbuf}}
\newcommand{\scbuf}{\mathsl{enc\_sendbuf}}
\newcommand{\rcbuf}{\mathsl{enc\_recvbuf}}
\newcommand{\IRecv}{\texttt{MPI\_IRecv} }
\newcommand{\Recv}{\texttt{MPI\_Recv} }
\newcommand{\ISend}{\texttt{MPI\_ISend} }
\newcommand{\Send}{\texttt{MPI\_Send} }

\newcommand{\commentt}[1]{\small {// #1}}
\newcommand{\FOR}{\textbf{for }}
\newcommand{\TO}{\textbf{to }}
\newcommand{\DO}{\textbf{do }}
\def\next{\:;\:}
\newcommand{\ind}{\hspace*{10pt}}
\newcommand{\flag}{\textsl{last}}
\newcommand{\IF}{\textbf{if }}
\newcommand{\ELSE}{\textbf{else }}
\newcommand{\THEN}{\textbf{then }}
\newcommand{\GCMEnc}{\mathsf{GCM}.\mathsf{Enc}}
\newcommand{\RETURN}{\textbf{return}}
\newcommand{\Header}{\textsl{Header}}
\newcommand*{\sys}{{\sffamily CryptMPI}\xspace}
\newcommand{\sysrm}{\textrm{CryptMPI}}
\newcommand{\Wait}{\texttt{MPI\_Wait} }
\newcommand{\Waitall}{\texttt{MPI\_Waitall} }
\newcommand{\EncAlltoall}{\texttt{Encrypted\_Alltoall} }
\newcommand{\EncAllgather}{\texttt{Encrypted\_Allgather} }
\newcommand{\EncAlltoallv}{\texttt{Encrypted\_Alltoallv} }
\newcommand{\EncIRecv}{\texttt{Encrypted\_IRecv} }
\newcommand{\EncBcast}{\texttt{Encrypted\_Bcast} }

 \newcommand{\CryptoPP}{\text{CryptoPP}}
\newcommand{\Exp}{\mathbf{E}}

\newcommand{\Alltoall}{\texttt{MPI\_Alltoall} }
\newcommand{\Alltoallv}{\texttt{MPI\_Alltoallv} }
\newcommand{\Bcast}{\texttt{MPI\_Bcast} }
\newcommand{\Allgather}{\texttt{MPI\_Allgather} }

\newcommand{\rank}{n}
\newcommand{\vecC}{\boldsymbol{C}}
\newcolumntype{C}[1]{>{\Centering}m{#1}}
\newcommand{\heading}[1]{\vspace{5pt}\noindent\underline{\textsc{#1}}}
\newcommand{\noskipheading}[1]{\noindent\underline{\textsc{#1}}}

\newcommand{\Tcomm}{T_{\mathrm{comm}}}
\newcommand{\acomm}{\alpha_{\mathrm{comm}}}
\newcommand{\bcomm}{\beta_{\mathrm{comm}}}
\newcommand{\Ted}{T_{\mathrm{enc}}}
\newcommand{\aed}{\alpha_{\mathrm{enc}}}
\newcommand{\bed}{\beta_{\mathrm{enc}}}
\newcommand{\Tall}{T_{\mathrm{ecom}}}
\newcommand{\aall}{\alpha_{\mathrm{ecom}}}
\newcommand{\ball}{\beta_{\mathrm{ecom}}}
\newcommand{\Red}{R_{\mathrm{enc}}}
\newcommand{\Rcomm}{R_{\mathrm{comm}}}
\newcommand{\Acomm}{A_{\mathrm{comm}}}
\newcommand{\Bcomm}{B_{\mathrm{comm}}}

\section{Related work} \label{sec:related}

\noskipheading{Prior systems of encrypted MPI.}
There have been a few proposed systems for adding encryption to MPI libraries, 
and some have even been
implemented~\cite{Ruan:2012:EMP:2197079.2197242,Balamurugan2014cmpich2,Maffina2013,RTSJ16,Shivaramakrishnan2014}. Existing systems, however, suffer
from notable security vulnerabilities, as we will elaborate below. 

First, privacy---the main goal of those systems---is seriously flawed because 
of the insecure crypto algorithms or the misuse of crypto algorithms.
For example, ES-MPICH2~\cite{Ruan:2012:EMP:2197079.2197242} is the first MPI
library that
integrates encryption to MPI communication, but its implementation is 
based on a weak encryption scheme, the Electronic Codebook (ECB) mode of 
operation. While ECB is still included in several standards, such as 
NIST SP 800-38A, ANSI X3.106, and ISO 8732, it has been known to be insecure~\cite[page 89]{Book:KL14}. 
For another example of an insecure choice of encryption, 
consider the system VAN-MPICH2~\cite{RTSJ16}
that relies on one-time pads for encryption. 
It however implements one-time pads as substrings of a big key~$K$. 
Thus when encrypting many large messages, 
it is likely that there are two messages $M_1$ and~$M_2$ whose one-time 
pads $L_1$ and $L_2$ are overlapping substrings of $K$, 
say the last 8 KB of $L_1$ is also the first 8 KB of $L_2$. 
In that case,  one can obtain the xor of $X_1$ and~$X_2$, 
where $X_1$ is the last 8 KB of $M_1$ and $X_2$ is the first 8 KB of~$M_2$. 
If $X_1$ and $X_2$ are English texts there are known methods to recover 
them from their xor value~\cite{CCS:OTP}. 

Next, no existing system provides meaningful data integrity. 
Some do suggest that 
integrity may be added via digital signatures~\cite{Ruan:2012:EMP:2197079.2197242,RTSJ16}, but this is impractical 
because all existing digital signature schemes are expensive. 
Some consider encrypting each message together with a checksum (obtained via a cryptographic hash function such as SHA-2)~\cite{Maffina2013}, 
but this approach does not provide integrity if one uses classical encryption schemes such as 
the Cipher Block Chaining (CBC) mode of encryption~\cite{C:HB01}. 
Others believe that encrypting data via the ECB mode also provides integrity~\cite{Ruan:2012:EMP:2197079.2197242,Balamurugan2014cmpich2}, 
but it is well-known that classical encryption schemes such as ECB or CBC provide no 
integrity at all~\cite[page 109]{Book:KL14}. 

We note that the insecurity of the systems above has never been realized in the literature. 
MPI communication therefore is in dire need of strong encryption that provides both privacy and integrity. 
In addition, in recent years, hardware support for 
efficient cryptographic operations, such as Intel's AES-NI instructions, 
has become ubiquitous. 
These advances are fully exploited by modern cryptographic libraries
to improve encryption speed.  
Yet 
there is currently a lack of understanding of how these libraries perform in
the MPI environment.  
Our paper fills this gap, giving (i) the first implementation that properly encrypts MPI communication
to provide genuine privacy and integrity, 
and (ii) a systematic benchmarking to investigate the overheads
of modern cryptographic libraries for MPI communication on contemporary clusters. 
Unlike prior work with insecure, ad hoc encryption schemes, 
our implementation is based on the Galois-Counter Mode (GCM) 
that \emph{provably} delivers both privacy and integrity~\cite{INDO:GCM04}. 

\heading{Performance modeling.}
Modeling the performance of communication systems has a long history; and 
many models have been developed. 
Most notable performance models include the Hockney
model \cite{Hockney94}, the LogP model \cite{Culler93}, 
the LogGP model \cite{Alexandrov95}, and the PLogP model \cite{Kielmann00}.
A good summary of these performance models is given
in \cite{Pjesivac05}. The Hockney model is widely used and has been
used to analyze MPI communication algorithms
\cite{Thakur05,Faraj05,Rabenseifner04,Faraj2008,PATARASUK2008809,Patarasuk:2009}. In this work, 
we show that the Hockney model can be used for the encrypted Ping Pong setting, 
and the recent max-rate model~\cite{gropp2016modeling} for concurrent one-to-one flows
can be used for multi-threading encryption. 
In hindsight, encrypted communication can be viewed as ordinary MPI communication over a private, authenticated channel, 
and thus it is understandable why conventional models for MPI communication continue to work well for encrypted settings.

\section{Background}



\subsection{Encryption Schemes}


\noskipheading{A bird's-eye view of encryption.}
A (symmetric) encryption scheme is a triple of algorithms $(\Gen, \Enc, \Dec)$. 
Initially, the sender and receiver somehow manage to share a secret key $K$ that is randomly generated by $\Gen$. 
Each time the sender wants to send a message $M$ to the receiver, 
she would encrypt $C \gets \Enc(K, M)$, 
and then send the ciphertext~$C$ in the clear. 
The receiver, upon receiving~$C$, will decrypt $M \gets \Dec(K, C)$. 
An  encryption scheme is commonly built on top of a blockcipher 
(such as AES and 3DES). 

Standard documents, such as NIST SP 800-38A~\cite{Dworkin:2001:SER:2206247} 
and 800-38D~\cite{Dworkin:2007:SRB:2206251}, specify several modes of encryption. 
Many of them, such as Electronic Codebook (ECB), 
Cipher Block Chaining (CBC), Counter (CTR), Galois/Counter Mode (GCM), 
and Counter with CBC-MAC (CCM), are well-known and widely used. 
However, these schemes are not 
equal in security and ease of correct use. 
The ECB mode, for example,  
is insecure~\cite{Book:KL14}. 
CBC and CTR modes provide only privacy, meaning that the adversary 
cannot even distinguish ciphertexts of its chosen messages
with those of uniformly random messages of the same length. 
They however do not provide data integrity in the sense that the adversary 
cannot modify ciphertexts without detection.\footnote{
Actually, the adversary can still replace a ciphertext with a prior one; 
this is known as
\emph{replay attack}. Here we do not consider such attacks. 
}
Among the standardized encryption schemes, only GCM and CCM satisfy 
both privacy and integrity, but GCM is the faster one~\cite{FSE:KR11}. 
Therefore,  in this paper, we will focus on GCM; 
one does not need to know technical details of GCM
to understand our paper. 


\heading{GCM Overview.}
According to NIST SP 800-38D,
the blockcipher for GCM must be AES, and correspondingly, the key length is either $128, 192,$ or $256$ bits. 
The longer key length means better security against 
brute-force attacks, but also slower speed. 
In this paper, we consider both 128-bit key (the most efficient version) 
and 256-bit key (the most secure one). 
AES-GCM is a highly efficient scheme~\cite{FSE:KR11}, provably meeting
both privacy and integrity~\cite{INDO:GCM04}. 
Due to its strength, AES-GCM appears in several network protocols, 
such as SSH, IPSec, and TLS.

\begin{figure}[t]
\centering
\includegraphics[width=0.45\textwidth]{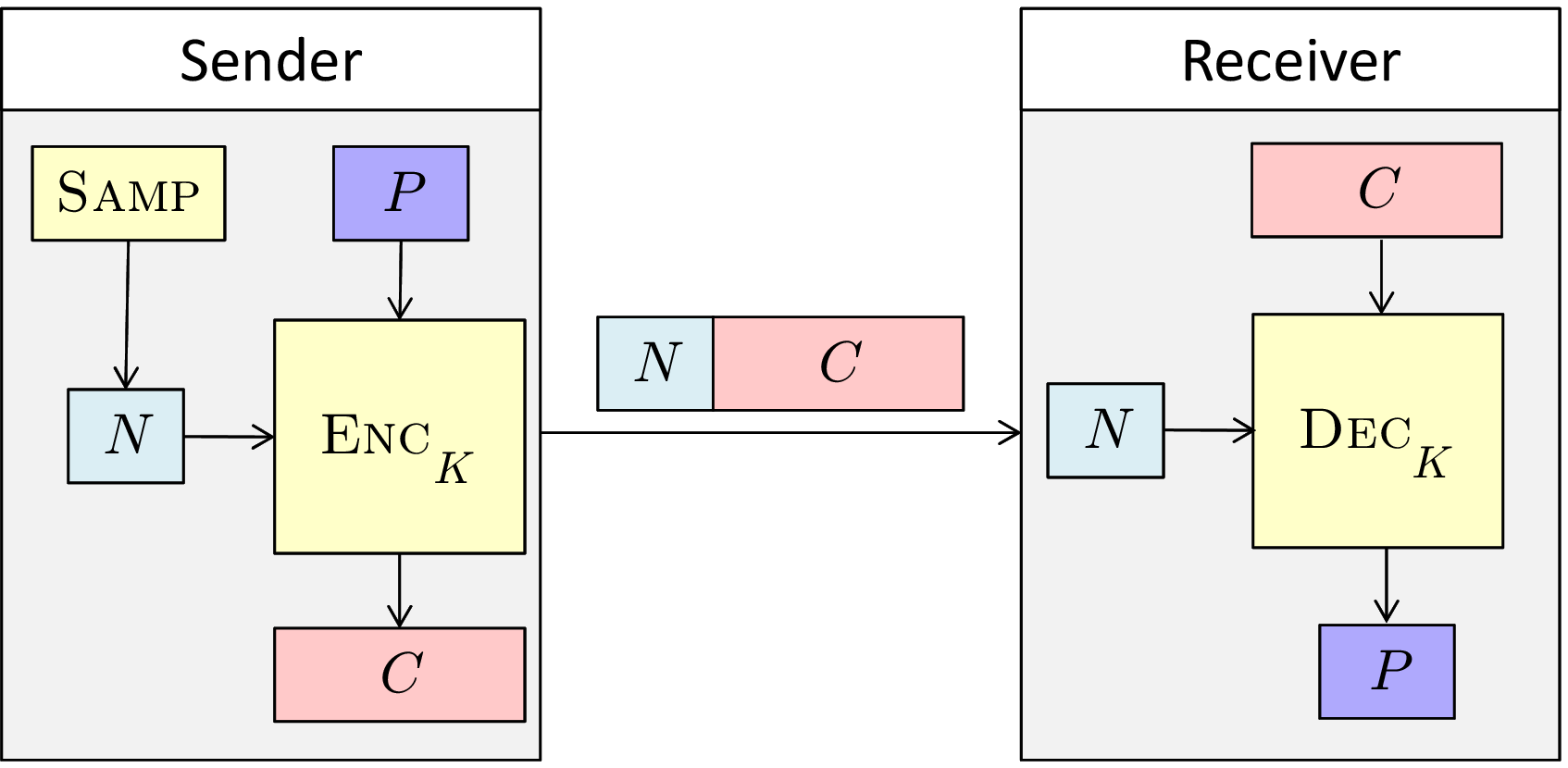}
\captionsetup{singlelinecheck=false}
\caption{Encrypted communication with AES-GCM.}
\label{fig:Enc}
\end{figure}

Syntactically, AES-GCM is a \emph{nonce-based} encryption scheme, 
meaning that to encrypt plaintext $P$, one needs to additionally provide a \emph{nonce} $N$, 
i.e., a public value that must appear at most once per key. 
The same nonce $N$ is required for decryption, and thus the sender needs to send both 
the nonce $N$ and the ciphertext~$C$ to the receiver. 
See \figref{fig:Enc} for an illustration of the encrypted communication via GCM. 
In AES-GCM, nonces are 12-byte long, 
and one often implements them via a counter, or pick them uniformly at random.  
In addition, each ciphertext is $16$-byte longer than the corresponding plaintext, 
as it includes a $16$-byte tag to determine whether the ciphertext is valid.

\subsection{Cryptographic Libraries}

In our implementation, we consider the following cryptographic libraries: OpenSSL~\cite{openssl}, 
BoringSSL~\cite{boringssl}, Libsodium~\cite{libsodium}, and $\CryptoPP$~\cite{cryptopp}. 
They are all in the public domain, are widely used, and have received 
substantial scrutiny from the security community. 

OpenSSL is one of the most popular cryptographic libraries, providing a 
widely used implementation of the Transport Layer Security (TLS) and 
Secure Sockets Layer (SSL) protocols. Due to its importance, 
there has been a long line of work in checking the security of OpenSSL, 
resulting in the discovery of several important vulnerabilities, 
such as the notorious Heartbleed bug~\cite{heartbleed}. 
As a popular commercial-grade toolkit, OpenSSL is used by many systems.
BoringSSL is Google's fork of OpenSSL, providing the SSL library in Chrome/Chromium and Android OS.

Libsodium is a well-known cryptographic library that aims for security and 
ease of correct use. 
It provides several benefits such as 
portability, cross-compilability, and API-compatibility, 
 and supports bindings 
for all common programming languages. As a result, Libsodium 
has been used in a number of applications, such as the cryptocurrency Zcash 
and Facebook's 
OpenR (a distributed platform for building autonomic network functions). It 
however only supports AES-GCM with 256-bit keys.

$\CryptoPP$ is another popular open-source cryptographic library for C++. 
It is widely adopted in both academic and commercial usage, including 
WinSSHD (an SSH server for Windows), 
Steam (a digital distribution platform purchasing and playing video games), and 
Microsoft SharePoint Workspace (a document collaboration software).

\section{MPI with encrypted communication}
\label{sec:libraries}
 
We developed two MPI libraries whose communication is encrypted via AES-GCM (for both $128$-bit and $256$-bit keys); 
one library is based on MPICH-3.2.1 for Ethernet and the other on MVAPICH2-2.3 for
InfiniBand. Specifically, encryption is added to the following MPI routines: 
\begin{itemize}
\item Point-to-point: $\Send$, $\Recv$, $\ISend$, $\IRecv$, $\Wait$, and $\Waitall$. 
\item Collective: $\Allgather$, $\Alltoall$, $\Alltoallv$,  and $\Bcast$. 
\end{itemize}
The underlying cryptographic library is user-selectable 
among OpenSSL, BoringSSL, Libsodium, and CryptoPP. 
With encryption incorporated at
the MPI layer, our prototypes can run
on top of any underlying network. 
As our main focus of this work is to 
benchmark the performance of encrypted MPI libraries, we did not 
implement a key distribution mechanism; this is left as a future work.  
In our experiments, the encryption key was hardcoded in the source code.

\RestyleAlgo{boxed}
\IncMargin{1em}
\begin{algorithm}{}
  \DontPrintSemicolon
  \SetKwFunction{alltoall}{MPI\_Alltoall}\SetKwFunction{nonce}{RAND\_bytes}
  \BlankLine
	\textbf{Input:} Two arrays $\sendbuf$ and $\recvbuf$, each of $n + 1$ elements
	that are $\ell$-byte long.\;
	\textbf{Parameter:} A key $K$\;
	\tcc{Create ciphertext buffers}
  Initialize two arrays $\scbuf$ and $\rcbuf$, each of $n + 1$ elements
	that are $(\ell+28)$-byte long. \;
  \For{$i\leftarrow 0$ \KwTo $\rank$}{	   
	    \tcc{Get a random 12-byte nonce}
			$N_i \gets \texttt{RAND\_bytes}(12)$\;
			\tcc{Encrypt via AES-GCM}
	    $C_i\gets \Enc(K, N_i, \sendbuf[i])$\;
			\tcc{Concatenate nonce and ctx}
			$\scbuf[i] \gets N_i \concat C_i$ \;
  }
  \BlankLine
	\alltoall{$\scbuf, \rcbuf$}\;
  \BlankLine
  \For{$i\leftarrow 0$ \KwTo $\rank$}{
	    \tcc{Parse to nonce/ciphertext}
	    $N^*_i \concat C^*_i \gets \rcbuf[i]$\;
			\tcc{Decrypt via AES-GCM}
			$\recvbuf[i] \gets \Dec(K, N^*_i, C^*_i)$ \;
  }
\vspace{2ex}
\caption{\EncAlltoall routine.}\label{algo:alltoall}
\end{algorithm}\DecMargin{1em}

To illustrate the high-level ideas of our implementation, 
consider the pseudocode of our \EncAlltoall routine
in Algorithm~\ref{algo:alltoall}. 
Within this code, we use $\texttt{RAND\_bytes}(s)$ to 
denote the sampling of a uniformly random $s$-byte string,
and $X \concat Y$ for the concatenation of two strings $X$ and $Y$. 
The encryption and decryption routines of AES-GCM are $\Enc$ and $\Dec$, respectively. 
Intuitively,  the ordinary $\Alltoall$ is used to send/receive just ciphertexts and their corresponding nonces. 
That is, one would need to encrypt the sending messages before calling $\Alltoall$---each message with a fresh random nonce, 
and then decrypt the receiving ciphertexts. 
If a sending message is $\ell$-byte long
then the corresponding data that $\Alltoall$ sends is $(\ell+28)$-byte long, 
since it consists of (i) a $12$-byte nonce, and (ii)
a ciphertext that is $16$-byte longer than its plaintext.

We note that although the pseudocode above seems straightforward, 
in an actual implementation, 
there are some low-level subtleties when one has to 
deal with non-blocking communication. 
For example, for $\EncIRecv$, 
our implementation performs decryption inside $\Wait$
to ensure the non-blocking property.

 \section{Experiments} 
\label{sec:perf}

We empirically evaluated the performance of our encrypted MPI libraries to 
(i) understand the encryption overheads in MPI settings, 
and (ii) determine the best cryptographic library to use with MPI. 
Below, we will first describe the system of our experiments, 
the benchmarks, and our methodology. 
Later, in \secref{sec:Ethernet} and \secref{sec:IB}, 
we will report the experiment results on Ethernet and Infiniband respectively.

\heading{System setup.}
The experiments were performed on a cluster with the following configuration.
The processors are Intel Xeon E5-2620 v4 with the base frequency of 2.10 GHz. Each node 
has 8 cores and 64GB DDR4 RAM and runs CentOS 7.6. Each node is equipped with
two types of network interface cards: 
a  10 Gigabit Ethernet card (Intel 82599ES SFI/SFP+)
and a 40 Gigabit InfiniBand QDR one (Mellanox  MT25408A0-FCC-QI ConnectX). 
Allocated nodes were chosen manually. For each experiment, the same node allocation was repeated
for all measurements.  All ping-pong results use two
processes on different  nodes.

We implemented 
our prototypes on top of MPICH-3.2.1 (for Ethernet) 
and MVAPICH2-2.3  (for Infiniband).  
The baseline and 
our encrypted MPI libraries were compiled 
with the standard set of MPICH and MVAPICH compilation flags and optimization level O2.  
In addition, we compiled all the cryptographic libraries 
(OpenSSL 1.1.1, BoringSSL, CryptoPP 7.0, and Libsodium 1.0.16)
separately using their default settings
and linked them with MPI libraries during the linking phase of MPICH and MVAPICH.

%


\heading{Benchmarks.} We consider the following suites of benchmarks: 

\begin{itemize}
\item Encryption-decryption: The encryption-decryption benchmark measures the
  encryption and decryption performance. For each data
  size, it measures the time for performing 500,000 times the simple
  encryption and then decryption of the data using a single thread. 
  
\item Ping-pong: This benchmark measures the uni-directional throughput when 
two processes communicate back and forth repeatedly using blocking send and receive. 
We  ran several experiments, each corresponding to a particular message size
within the range from 1B to 2MB. 
In each experiment measurement, 
the two processes send messages of the designated size back and forth 10,000 times
if the message size is less than 1MB, and 1,000 times otherwise. 
For encrypted communication, each message results in an additional 28-byte overhead, 
as we need to send a 12-byte nonce and a 16-byte tag per ciphertext. 
Those bytes 
are excluded in the throughput calculation. 

\item OSU micro-benchmark 5.4.4~\cite{OSUBM}: We used the Multiple Pair Bandwidth Test
benchmark in OSU suite to measure aggregate uni-directional throughput when multiple senders in one node
communicate  with their corresponding receivers in another node, via non-blocking send and receive. 
In each experiment measurement, the sender iterates 100 times; 
in each iteration, it sends 64 messages of the designated size to the receiver and wait for the replies
before moving to the next iteration. 
Again, we excluded the 28-byte overhead per message in calculating the throughput.

\smallskip 

We also used OSU suite to measure performance of collective communication routines. 
Each experiment measurement consists of 100 iterations.

\item NAS parallel benchmarks~\cite{Bailey:1991:NPB:2748645.2748648}: 
To measure performance of (encrypted) MPI in applications, 
we used the BT, CG, FT, IS, LU,
MG, and SP in the NAS parallel benchmarks; 
all experiments used Class C size. 
\end{itemize}

\heading{Benchmark methodology.}
For ping-pong, OSU benchmarks, and NAS benchmarks,
we first ran each experiment at
least 20 times, up to 100 times, 
until the standard deviation was within 5\% of the arithmetic mean.
If after 100 measurements, the standard deviation  was still too big then
we would keep running the experiment 
 until the 99\% confidence interval 
was within 5\% of the mean. The variability for encryption and decryption is
much smaller. Hence, each result for the encryption-decryption benchmark
is obtained by running the benchmark at least 5 times until the standard
deviation was within 5\% of the arithmetic mean. 



To evaluate the scalability of our implementation, 
we used four different settings (e.g. 4 rank/4 node, 16 rank/4 node, 
16 rank/8 node and 64 rank/8 node) for OSU and NAS benchmarks. 

\heading{What we report.}
In our experiments, BoringSSL and OpenSSL delivered very similar performance.
This is not surprising, since BoringSSL is a fork of OpenSSL. 
In addition, the benchmarks yielded the same trends for both 128-bit and 256-bit keys. 
We therefore only report the results of BoringSSL (256-bit key), Libsodium, and CryptoPP (256-bit key).

\subsection{Ethernet Results} \label{sec:Ethernet}

\begin{figure}[t]
	\centering
	\includegraphics[width=0.45\textwidth]{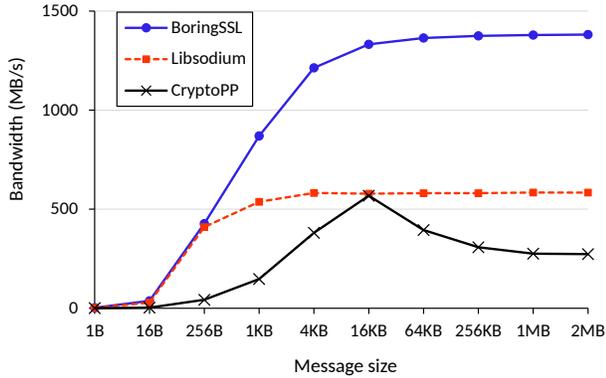}
	        \caption{Encryption-decryption throughput of AES-GCM-256, compiled with
                the gcc 4.8.5.}
	\label{fig:enc_dec_gcc_throughput}
\end{figure}

\noskipheading{Encryption-decryption.} Before we get into the details of the
communication benchmark results, 
it is instructive to understand the performance of AES-GCM, 
since it helps us to have a better understanding of the performance
of the encrypted MPI libraries. 
The average throughputs of AES-GCM-256 with different data sizes
are shown in \figref{fig:enc_dec_gcc_throughput}. It is clear that
different encryption libraries have very different encryption and
decryption performance. There are two ways to interpret the results
here. First, one can view this as the convergence of the ping-pong performance
when the network speed becomes much faster than encryption and descryption. 
Also, since for AES-GCM, the encryption and decryption speed
is roughly the same, the reported performance here is a half of the
encryption throughput (that is also decryption throughput).  
Thus we can predict that for most experiments, among the three
encrypted MPI libraries, BoringSSL is the best, and then Libsodium,
and finally CryptoPP.

\begin{figure}[t]
	\centering
		\includegraphics[width=0.45\textwidth]{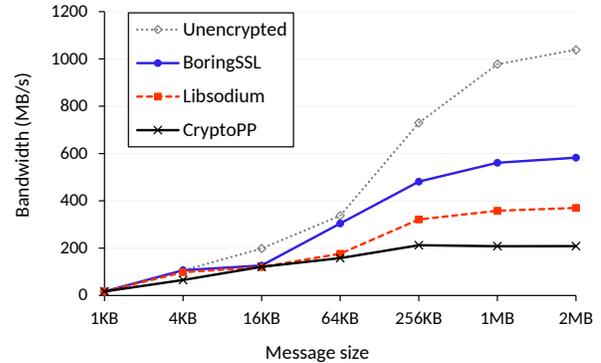}
	\caption{Average unidirectional ping-pong throughput  with 256-bit encryption key on Ethernet, 
	for medium and large messages.}
	\label{fig:pingpong_ethernet}
	\vspace{-1ex}
\end{figure}

\heading{Ping-pong.} The ping-pong performance of the baseline and the encrypted MPI libraries is
shown in Table~\ref{tab:pingpong_ethernet} for small messages, 
and 
in \figref{fig:pingpong_ethernet} for medium and large messages. 
For 1KB data,  
BoringSSL appears to slightly outperform the baseline, 
but recall that we are reporting 
the mean values with 5\% deviation,
so this only means that BoringSSL has close performance to 
the~baseline.

\bigskip 

\begin{minipage}[t]{\linewidth}
\centering 
\captionsetup{justification=centering, labelsep=newline}
\captionof{table}{Average Unidirectional Ping-Pong Throughput  (MB/s) for Small Messages, with 256-bit Encryption Key on Ethernet.} 
\label{tab:pingpong_ethernet}
\begin{tabular}{ C{.9in}  C{.35in} *3{C{.35in}}}\toprule[1.25pt]
 &  \textbf{1B} & \textbf{16B} & \textbf{256B} & \textbf{1KB}  \\ \midrule
\textbf{Unencrypted} & 0.050 & 0.83 & 7.01 & 17.03  \\ \midrule
\textbf{BoringSSL} & 0.045 & 0.78 & 6.62 & 17.05  \\ \midrule
\textbf{Libsodium} & 0.046 & 0.79 & 6.62 & 17.02  \\ \midrule
\textbf{CryptoPP}  & 0.029  &  0.48 &  6.85 & 17.02 \\
\bottomrule[1.25pt]
\end{tabular}
\bigskip
\end{minipage}

For large messages, say 2MB ones, encrypted MPI libraries have poor performance compared to the baseline: 
even the fastest BoringSSL yields 78.3\% overhead, 
and CryptoPP's overhead is much worse, nearly 400\%.
These performance results can be explained as follows. 

\begin{itemize}
\item The running time of an encrypted MPI library consists of (i) the encryption-decryption cost, 
and (ii) the underlying MPI communications, which roughly corresponds to the baseline performance. 

\item For BoringSSL, on 2MB messages, the encryption-decryption throughput of AES-GCM-256 (1381 MB/s) is about 1.32
times that of the ping-pong throughput of the baseline (1038 MB/s). 
Estimatedly, BoringSSL's ping-pong time
would be roughly $\frac{1 + 1.32}{1.32} \approx 1.76$ times slower than that of the baseline. 
This is consistent with the reported 78.3\% overhead above. 

\item For CryptoPP, on 2MB messages, the encryption-decryption throughput of AES-GCM-256 (273 MB/s) much worse, just 
around 26\% of the ping-pong performance of the baseline (1038 MB/s). 
One thus can estimate that CryptoPP's ping-pong time
would be about $\frac{1 + 0.26}{0.26} \approx 4.84$ times slower than that of the baseline. 
This is again consistent with the reported 400\% overhead above. 
\end{itemize}

For small messages, encrypted MPI libraries often perform reasonably well, 
since the encryption-decryption throughput of AES-GCM-256 is quite higher than 
the ping-pong throughput of the baseline. 
For example, for 256-byte messages,  the encryption-decryption throughput of Libsodium is 409.67 MB/s, 
much higher than the 7.01 MB/s baseline ping-pong throughput. 
Consequently, Libsodium has just 5.89\% overhead for 256-byte messages.

\begin{figure}[t]	
\centering
\subfloat[$1$B-messages]{
  \includegraphics[width=0.4\textwidth]{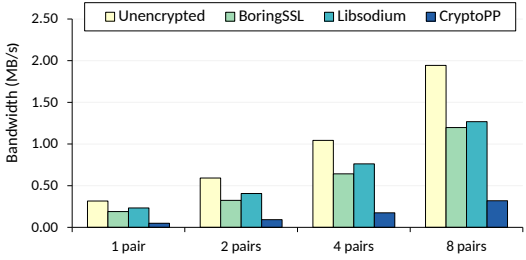}%
}

\subfloat[$16$KB-messages]{
  \includegraphics[width=0.4\textwidth]{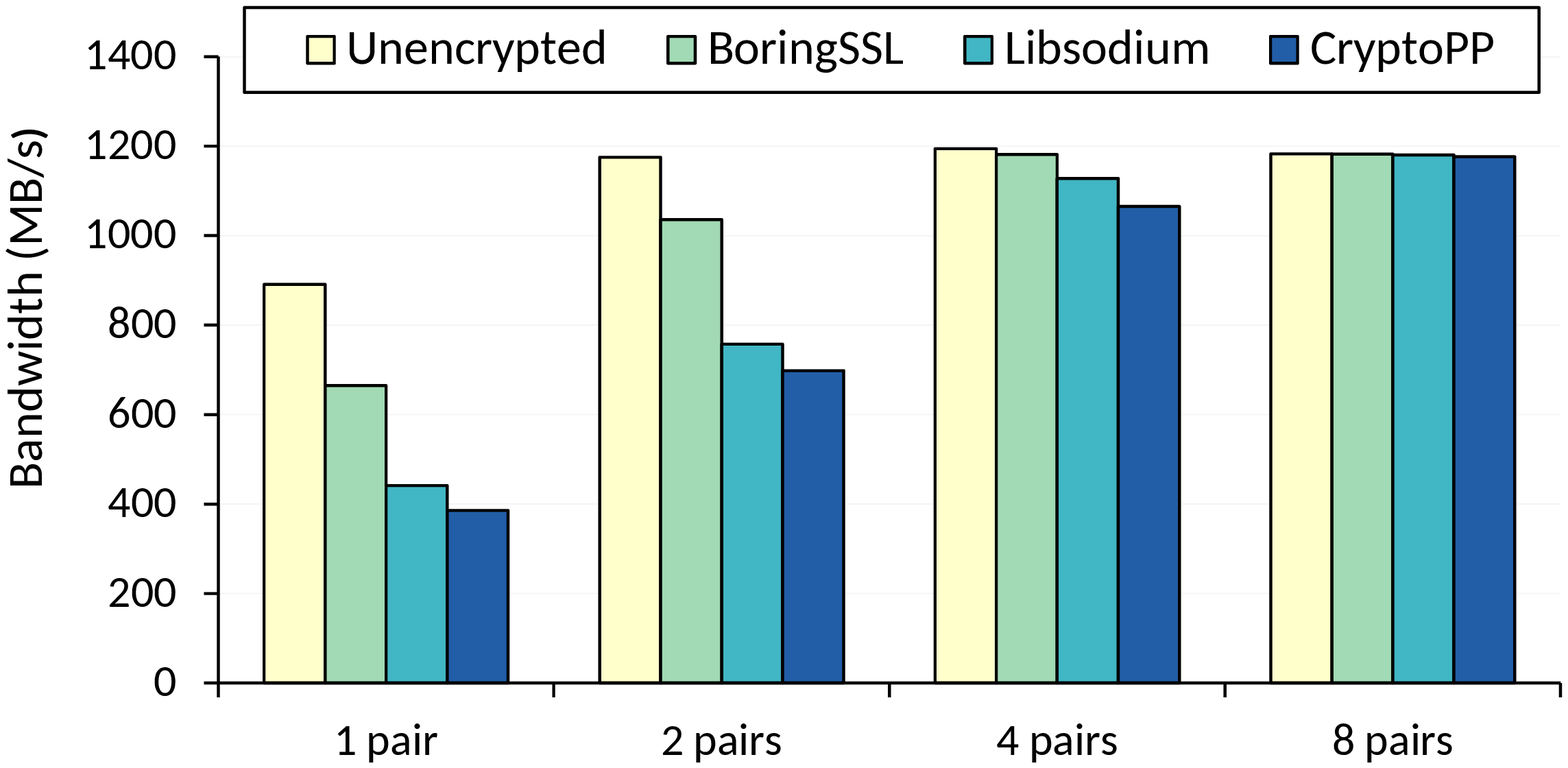}%
}

\subfloat[2MB-messages]{
  \includegraphics[width=0.4\textwidth]{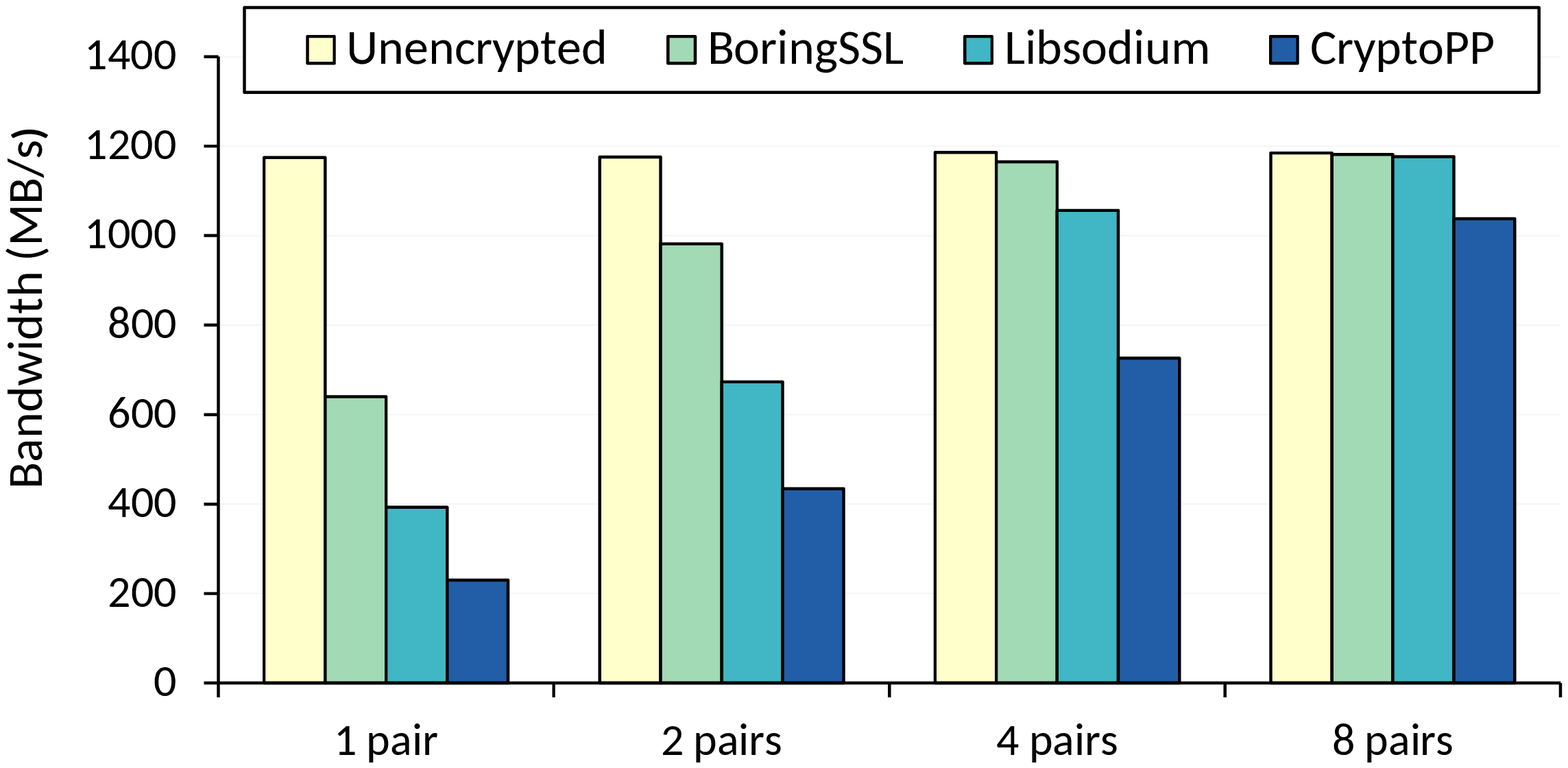}%
}
\captionsetup{singlelinecheck=false}
\caption{OSU Multiple-Pair average throughput on Ethernet. 
}
	\label{fig:multipair_ethernet_1B}
		\label{fig:multipair_ethernet_2M}
\vspace{-1.5ex}
\end{figure}

\heading{OSU Multiple-Pair Bandwidth.}
The Multiple-Pair performance of the baseline and the encrypted MPI libraries, 
for 1B, 
16KB, and 2MB messages, is shown in 
Figure~\ref{fig:multipair_ethernet_1B}. 

For medium and large messages, as the number of pairs increases, the relative performance of the encrypted MPI libraries becomes much better, 
because (i) the network bandwidth remains the same, yet the computational power doubles, 
and (ii) encryption/decryption can overlap with MPI communications. 
When there is just a single pair, even BoringSSL cannot encrypt fast enough to keep up with the network speed. 
However, when there are 8 pairs, even CryptoPP can reach the baseline performance, 
for 16KB messages. 
These results suggest that (1) the overhead for a single communication flow may be significant, 
but (2) in modern multi-core machines, when multiple
flows happen concurrently, the performance of encrypted MPI libraries may be on par with the baseline.

For small messages, the situation is different, because the network bandwidth is not fully used. 
As shown in \figref{fig:multipair_ethernet_1B}, for 1B-data, the baseline throughput keeps increasing as the number of pairs increases. 
In contrast, 
for medium and large messages, the baseline throughput is saturated when there are just two pairs of senders and receivers. 
Consequently, even when there are 8 pairs, BoringSSL still incurs 61.67\% overhead, 
and CryptoPP is far worse, resulting in 506.25\% overhead.

\heading{Collective communication.}
The average running time of $\EncBcast$
and 
$\EncAlltoall$, 
for the 64-rank and 8-node setting,  is shown in 
Tables~\ref{tab:bcast_ethernet}
and \ref{tab:alltoall_ethernet}, respectively.


\begin{figure}[t]

\begin{minipage}{\linewidth}
\centering
\captionsetup{justification=centering, labelsep=newline}
\captionof{table}{Average Timing of $\EncBcast$ ($\mu$s),   with 256-bit 
Key on Ethernet.} 
\label{tab:bcast_ethernet} 
\begin{tabular}[t]{ C{1in}  *2{C{.45in}} C{.8in}}\toprule[1.25pt]
 &  \textbf{1B} & \textbf{16KB} & \textbf{4MB}   \\ \midrule
\textbf{Unencrypted} & 31.15 & 231.75 & 9,594.75    \\ \midrule
\textbf{BoringSSL} &   37.15 & 246.17 & 13,892.74\\ \midrule
\textbf{Libsodium} &  35.54 & 264.37 & 18,322.19\\ \midrule
\textbf{CryptoPP}  &  54.97 & 278.65 & 29,301.96\\
\bottomrule[1.25pt]
\end{tabular}

\medskip
\end{minipage}
\end{figure}

\begin{figure}[t]
\begin{minipage}{\linewidth}
\centering
\captionsetup{justification=centering, labelsep=newline}
\captionof{table}{Average Timing of $\EncAlltoall$ ($\mu$s),  with 256-bit 
Key on Ethernet.} 
\label{tab:alltoall_ethernet} 
\begin{tabular}{ C{0.9in}  *2{C{.45in}} C{.8in}}\toprule[1.25pt]
 &  \textbf{1B} & \textbf{16KB} & \textbf{4MB}   \\ \midrule
\textbf{Unencrypted} &   159.13 & 6,562.82 & 1,966,299.47 \\ \midrule
\textbf{BoringSSL} &  329.60 & 7,691.08 & 2,210,546.32\\ \midrule
\textbf{Libsodium} &  452.76 & 8,937.74 & 2,535,104.93\\ \midrule
\textbf{CryptoPP}  &   1,221.98 & 9,462.90 & 3,297,402.93\\
\bottomrule[1.25pt]
\end{tabular}
\medskip
\end{minipage}
\end{figure}

\medskip

\begin{figure}[t]
\centering
\subfloat[$\EncBcast$]{
  \includegraphics[width=0.40\textwidth]{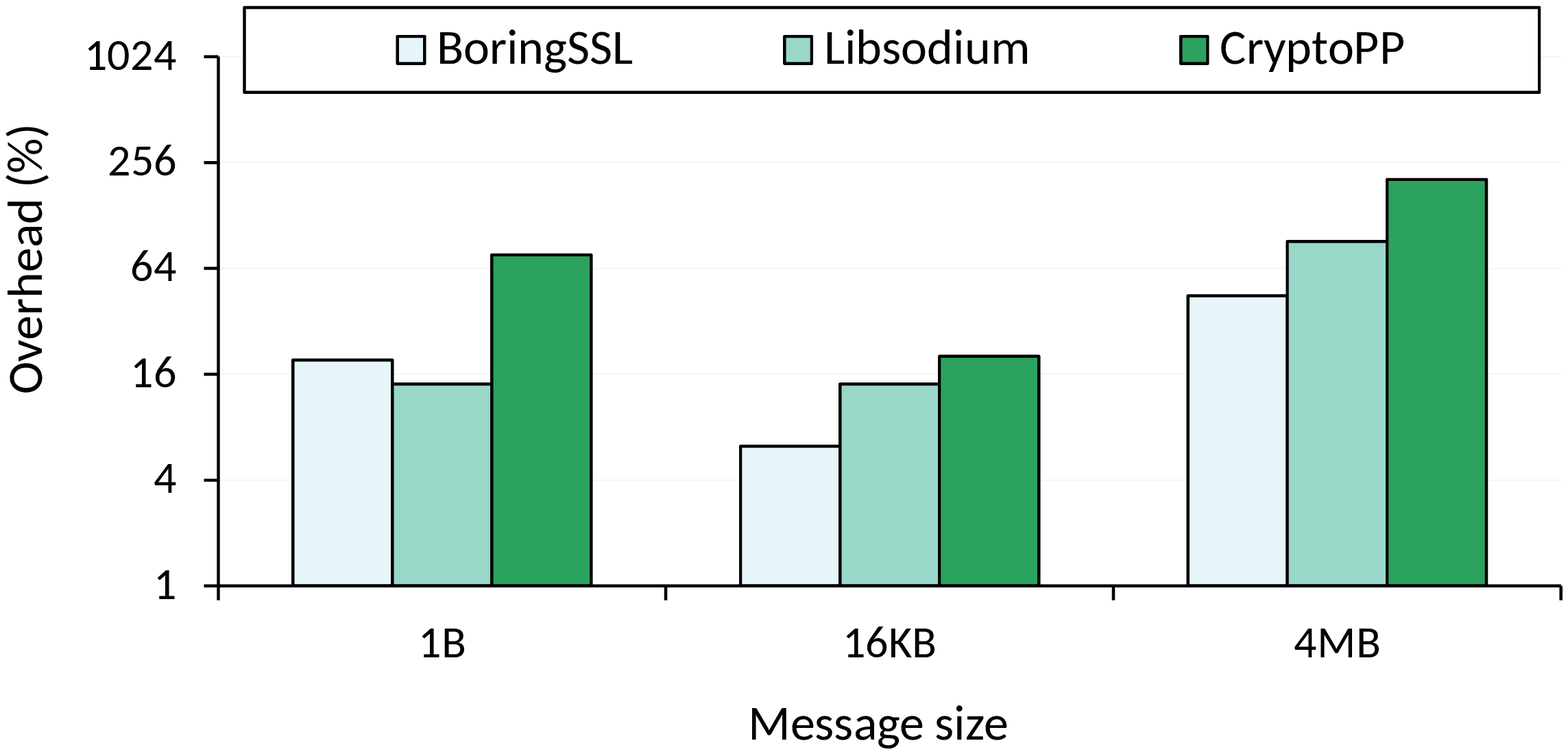}%
}
\\
\subfloat[$\EncAlltoall$]{
  \includegraphics[width=0.40\textwidth]{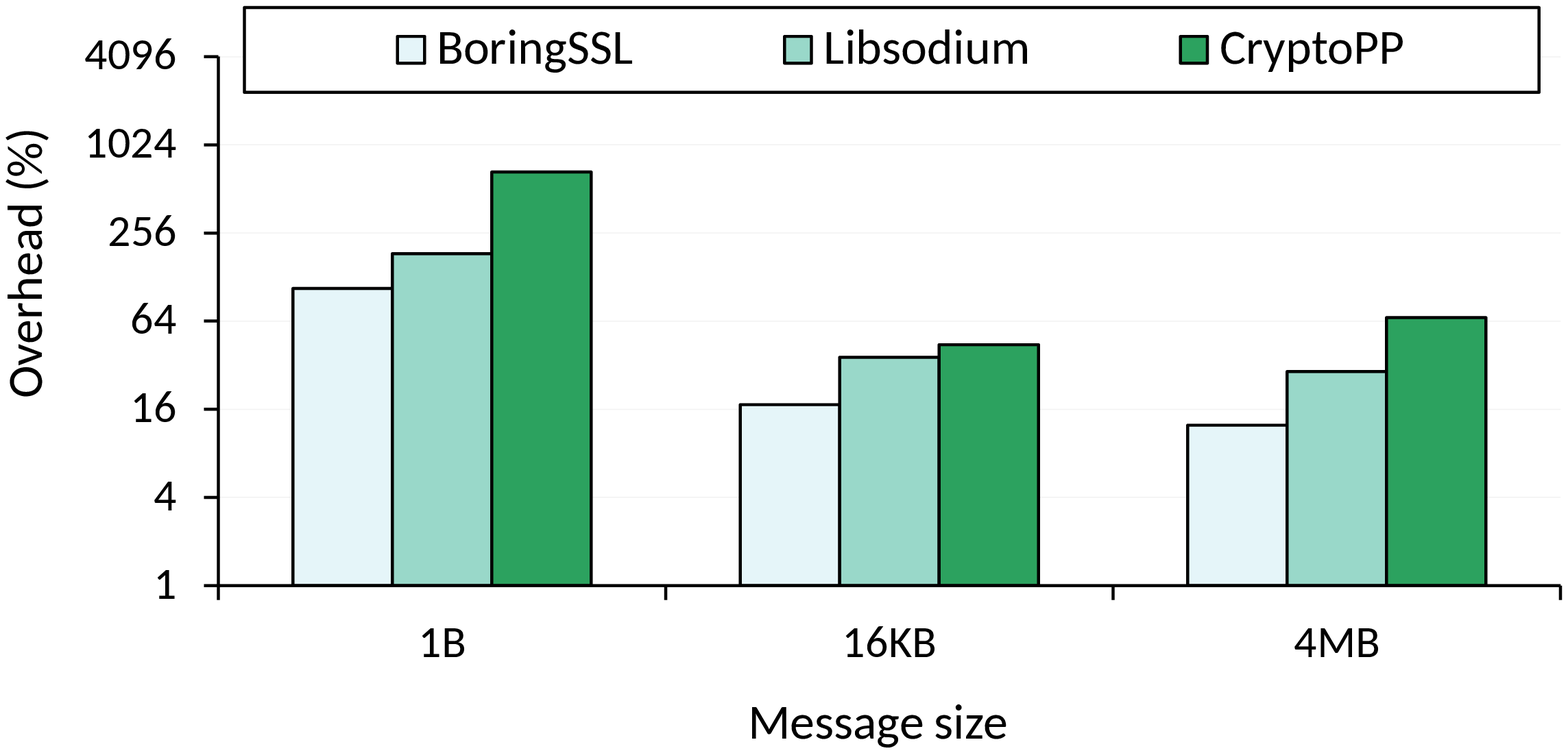}%
}
\caption{Encryption overhead (256-bit key), drawn in log scale, for 
$\EncBcast$ (left) and $\EncAlltoall$ (right) on Ethernet. 
}
	\label{fig:Bcast_Ethernet}
		\label{fig:Alltoall_Ethernet}
		\vspace{-1.5ex}
\end{figure}

To understand the performance of $\EncBcast$, 
recall that each encrypted broadcast consists of an ordinary $\Bcast$ and an encryption and a decryption per node. 
Hence the encryption overhead of the three encrypted MPI libraries, illustrated in \figref{fig:Bcast_Ethernet}, 
 loosely mirrors their encryption-decryption throughput
of \figref{fig:enc_dec_gcc_throughput}. 
\begin{itemize}
\item For example, for large messages (say 2MB), the encryption-decryption
throughput of BoringSSL (1381 MB/s) is around 2.37 times that of Libsodium (583 MB/s). 
On the other hand, the encryption overhead in $\EncBcast$ of BoringSSL (44.8\%)
is 2.03 times smaller than that of Libsodium (90.96\%), 
approximating the ratio 2.37 above. 
\item As another example, for BoringSSL, the encryption-decryption throughput for 2MB messages
is about the same as that for 16KB messages. 
Thus one would expect the encryption cost for 4MB messages in $\EncBcast$ would be about $\frac{4\text{MB}}{16\text{KB}} = 256$
times that for 16KB messages. 
Indeed, for 4MB messages, BoringSSL spends about 4,298 $\mu$s on encryption/decryption, 
which is about 298 times its encryption/decryption time for 16KB messages (14.42 $\mu$s). 
\end{itemize}

The trend of $\EncAlltoall$, illustrated in \figref{fig:Alltoall_Ethernet}, 
is similar.  
\begin{itemize}
\item For example, for 16KB messages, the encryption-decryption throughput of BoringSSL (1332 MB/s)
is about $2.35$ times that of CryptoPP (568 MB/s). 
The encryption overhead of BoringSSL in $\EncAlltoall$ (17.19\%)
is 2.57 times smaller than that of CryptoPP (44.19\%).  
\item As another example, for CryptoPP, the encryption-decryption throughput for 2MB messages
(273 MB/s) is about a half of that for 16 KB messages (568 MB/s). 
Thus one would expect the encryption cost for 4MB messages in $\EncAlltoall$ would be about $\frac{4\text{MB}}{16\text{KB}} \cdot 2 = 512$
times that for 16KB messages. 
Indeed, for 4MB messages, CryptoPP spends about  1,331,103 $\mu$s on encryption/decryption, 
which is about 459 times its encryption/decryption time for 16KB messages (2900~$\mu$s).
\end{itemize}

\begin{figure*}
\begin{minipage}{\linewidth}
\centering
\captionsetup{justification=centering, labelsep=newline}
\captionof{table}{Average Running Time (Seconds) of NAS Parallel Benchmarks, Class C, 64-rank \\ and 8-node, on Ethernet.}
\label{tab:NAS_Ethernet}
\begin{tabular}{C{0.16\linewidth}*7{C{0.067\linewidth}}}
\toprule[1.25pt]
 & \textbf{CG} & \textbf{FT} & \textbf{MG} & \textbf{LU} & \textbf{BT} & \textbf{SP} & \textbf{IS}  \\ \midrule
\textbf{Unencrypted} & 7.01 &  12.04 & 2.55 & 18.04 & 22.83 & 21.99 & 4.06  \\ \midrule
\textbf{BoringSSL} & 8.55 &  12.81 & 3.01 & 19.05 & 27.40 & 24.46 & 4.52 \\ \midrule
\textbf{Libsodium} &  9.62 & 13.67 & 3.09 & 19.48 & 28.70 & 26.30 & 4.71    \\ \midrule
\textbf{CryptoPP} & 11.67 & 15.53 & 3.33 & 23.13 & 29.52 & 27.37 & 4.83  \\
\bottomrule[1.25pt]
\end{tabular}
\end{minipage}
\end{figure*}

\medskip 

\heading{NAS benchmarks.}
To understand encryption overheads in a more realistic 
setting, we evaluated the encrypted MPI libraries under NAS parallel benchmarks.
The results are shown in Table~\ref{tab:NAS_Ethernet}. 
Overall, BoringSSL's total running time is 99.81 seconds, whereas the baseline's running time is 88.52 seconds, 
and thus BoringSSL's overhead is 12.75\%.\footnote{
Conventionally, one would compute BoringSSL's overhead of each benchmark (BT, CG, FT, etc)
and then report the average of them as BoringSSL's average overhead. 
However, as pointed out by several papers~\cite{fleming1986not,hoefler2015scientific}, 
averaging over ratios is \emph{meaningless}.
Following the recommendation of those papers, 
here we  instead derived BoringSSL's overhead from its total timing of all NAS benchmarks
and that of the baseline. 
}
Likewise, Libsodium's and CryptoPP's overhead are 19.25\% and 30.33\% respectively. 
These results again support our thesis 
that encryption overheads may not be prohibitive
for realistic scenarios where there are multiple concurrence communication flows.

\subsection{Infiniband Results} \label{sec:IB}

\begin{figure}[t]
	\centering
		\includegraphics[width=0.45\textwidth]{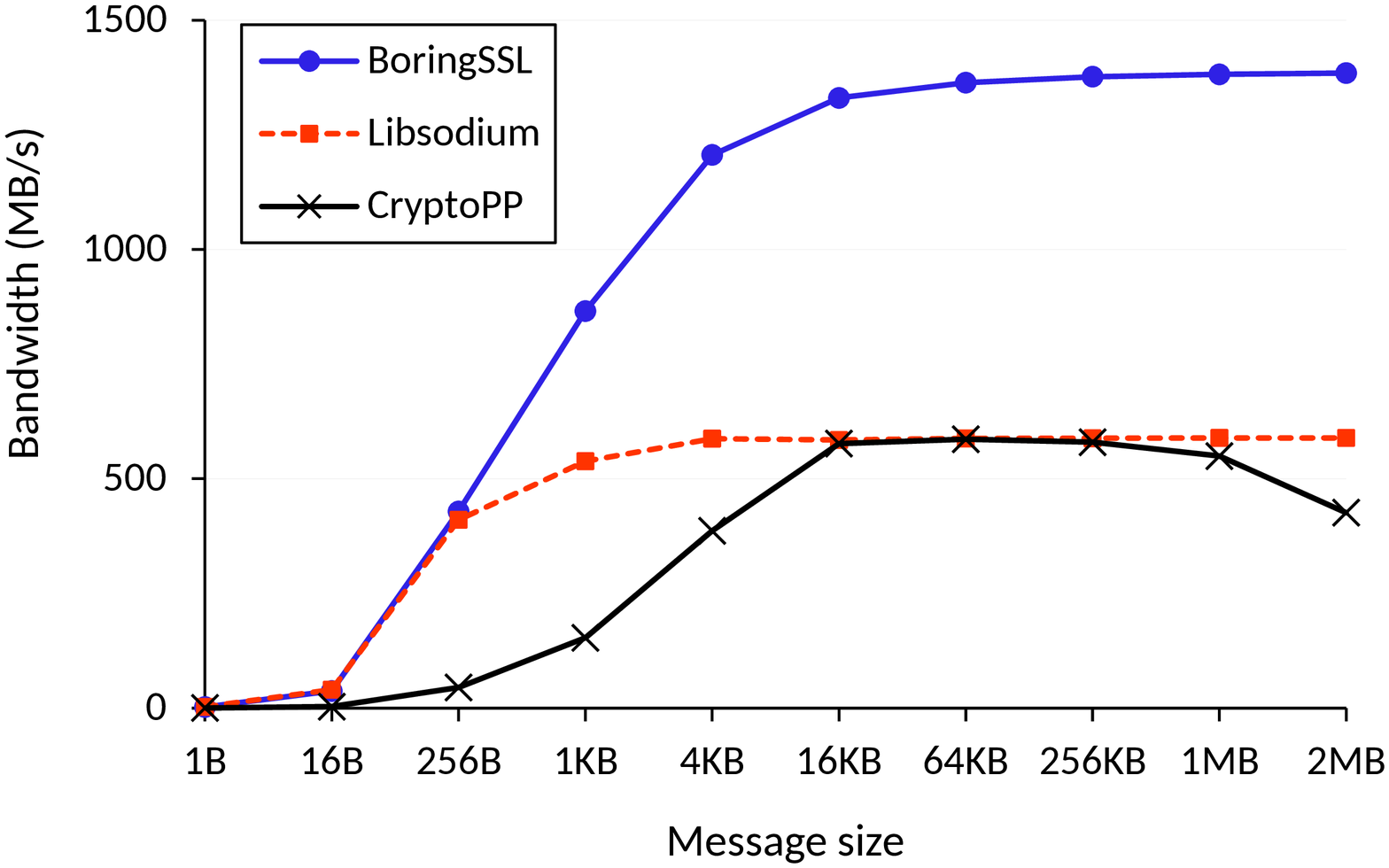}
	\caption{Encryption-decryption throughput of AES-GCM-256, compiled under MVAPICH2-2.3.}
	\label{fig:enc_dec_mvapich}
\end{figure}

\noskipheading{Encryption-decryption.}
It turns out that 
the MVAPICH2-2.3 compiler, even with the same O2 flag, 
results in higher encryption-decryption performance than 
the gcc 4.8.5 compiler for some libraries. 
\figref{fig:enc_dec_mvapich} shows the average encryption-decryption
throughput of AES-GCM-256 code compiled by the MVAPICH compiler. 
In particular, the performance of CryptoPP for message size greater than 64KB
is dramatically improved. 
It seems natural to predict that while CryptoPP is still the last among the three encrypted MPI libraries, 
for large messages, its performance will be close to that of Libsodium.

\begin{figure}[t]
	\centering
		\includegraphics[width=0.45\textwidth]{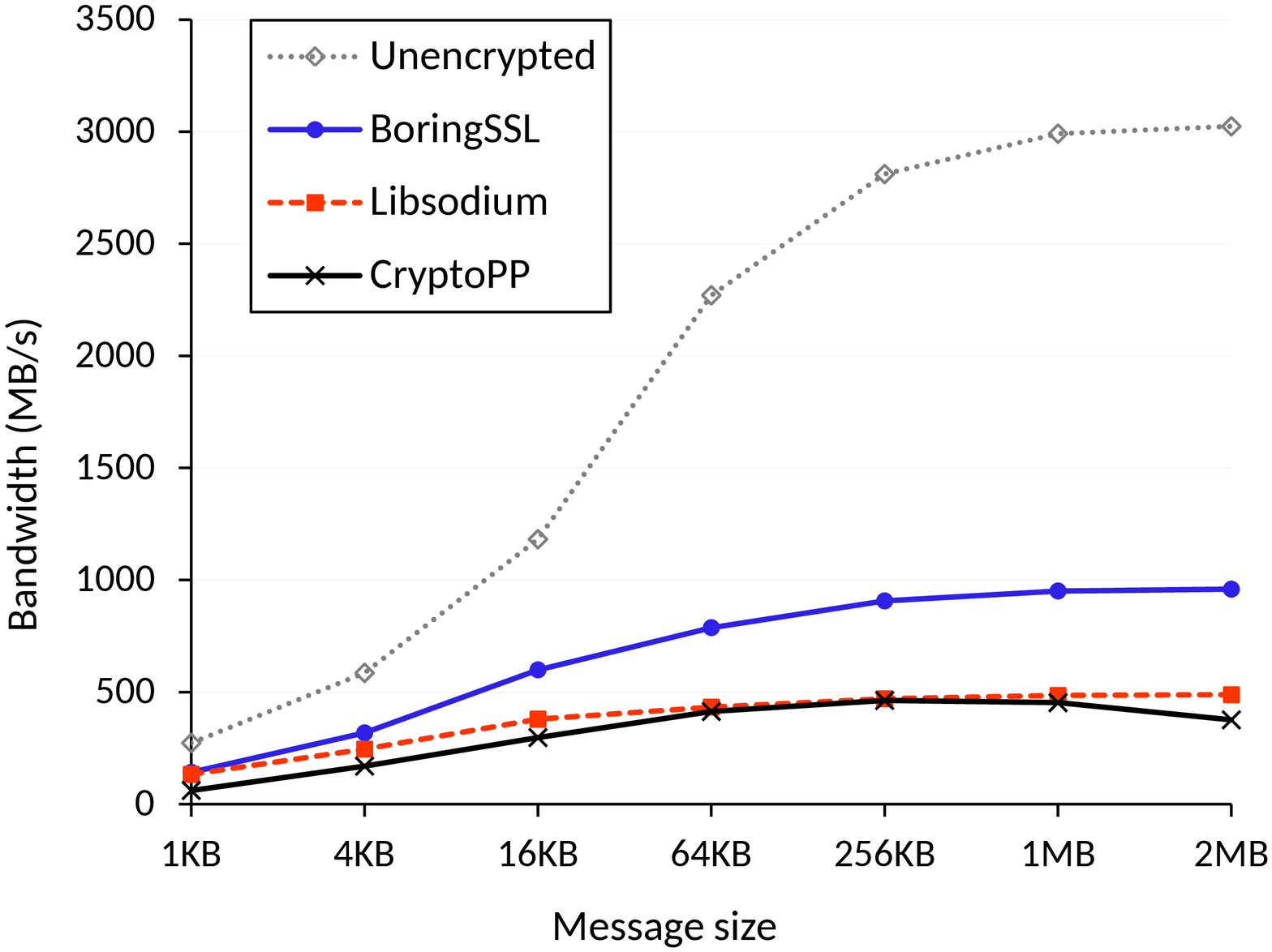}
	\caption{Unidirectional ping-pong throughput with 256-bit encryption key on Infiniband, 
	for medium and large messages.}
	\label{fig:pingpong_infiniband}
	\vspace{1ex}
\end{figure}

\heading{Ping-pong.} The ping-pong performance of the baseline and the encrypted MPI libraries is
shown in Table~\ref{tab:pingpong_infiniband} for small messages, 
and illustrated in \figref{fig:pingpong_infiniband} for medium and large messages.

Again, for large messages, the performance of the encrypted MPI libraries
is much lower than 
that of the baseline, but the situation is much worse than the
Ethernet setting. 
For example, for 2MB messages, even BoringSSL results in a 215.2\% overhead.
With InfiniBand, the baseline ping-pong throughput is significantly
higher than that with Ethernet while the encryption-decryption
throughput remains the same: the encryption-decryption throughput of
AES-GCM-256 is much lower than the ping-pong throughput of the baseline. 
For example, for 2MB messages, the encryption-decryption throughput
of AES-GCM-256  in BoringSSL (1384 MB/s) is just around 46\% of
the baseline ping-pong throughput (3023 MB/s), 
and thus estimatedly, BoringSSL's ping-pong time would be
about $\frac{1 + 0.46}{0.46} \approx 3.17$ times slower than
that of the baseline. This is consistent with the reported
215.2\% overhead above.

\bigskip 
\begin{figure}[!t]
\begin{minipage}{\linewidth}
\centering
\captionsetup{justification=centering, labelsep=newline}
\captionof{table}{Average Unidirectional Ping-Pong Throughput  (MB/s) for Small Messages,  with \\256-bit Encryption Key  on Infiniband.} \label{tab:pingpong_infiniband} 
\begin{tabular}{ C{1.25in}  C{.35in} *3{C{.35in}}}\toprule[1.25pt]
 &  \textbf{1B} & \textbf{16B} & \textbf{256B} & \textbf{1KB}  \\ \midrule
\textbf{Unencrypted} & 0.57 & 9.61 & 82.34 & 272.84  \\ \midrule
\textbf{BoringSSL} & 0.22 & 4.02 & 45.51 & 142.23  \\ \midrule
\textbf{Libsodium} & 0.27 & 4.86 &  50.66 & 133.06  \\ \midrule
\textbf{CryptoPP}  & 0.05  &  0.98 &  17.27 & 61.08 \\
\bottomrule[1.25pt]
\end{tabular}
\medskip
\end{minipage}
\end{figure}
\smallskip 

For small messages, the situation is somewhat better, but even BoringSSL would yield 
poor performance. For example, for 256-byte messages, BoringSSL has a 80.93\% overhead.

\begin{figure}[t]	
\centering
\subfloat[$1$B-messages]{
  \includegraphics[width=0.4\textwidth]{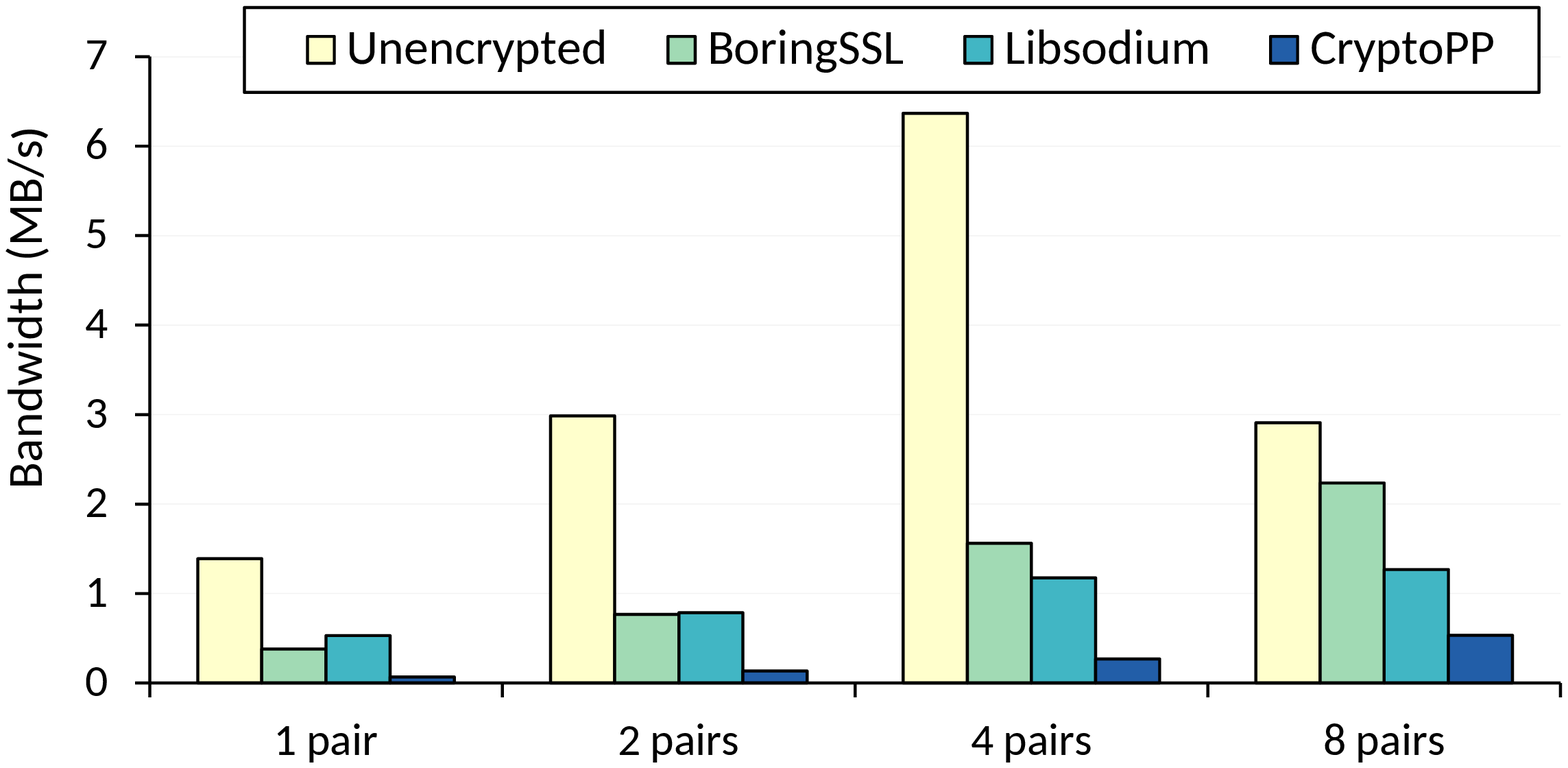}%
}
\\
\subfloat[$16$KB-messages]{
  \includegraphics[width=0.4\textwidth]{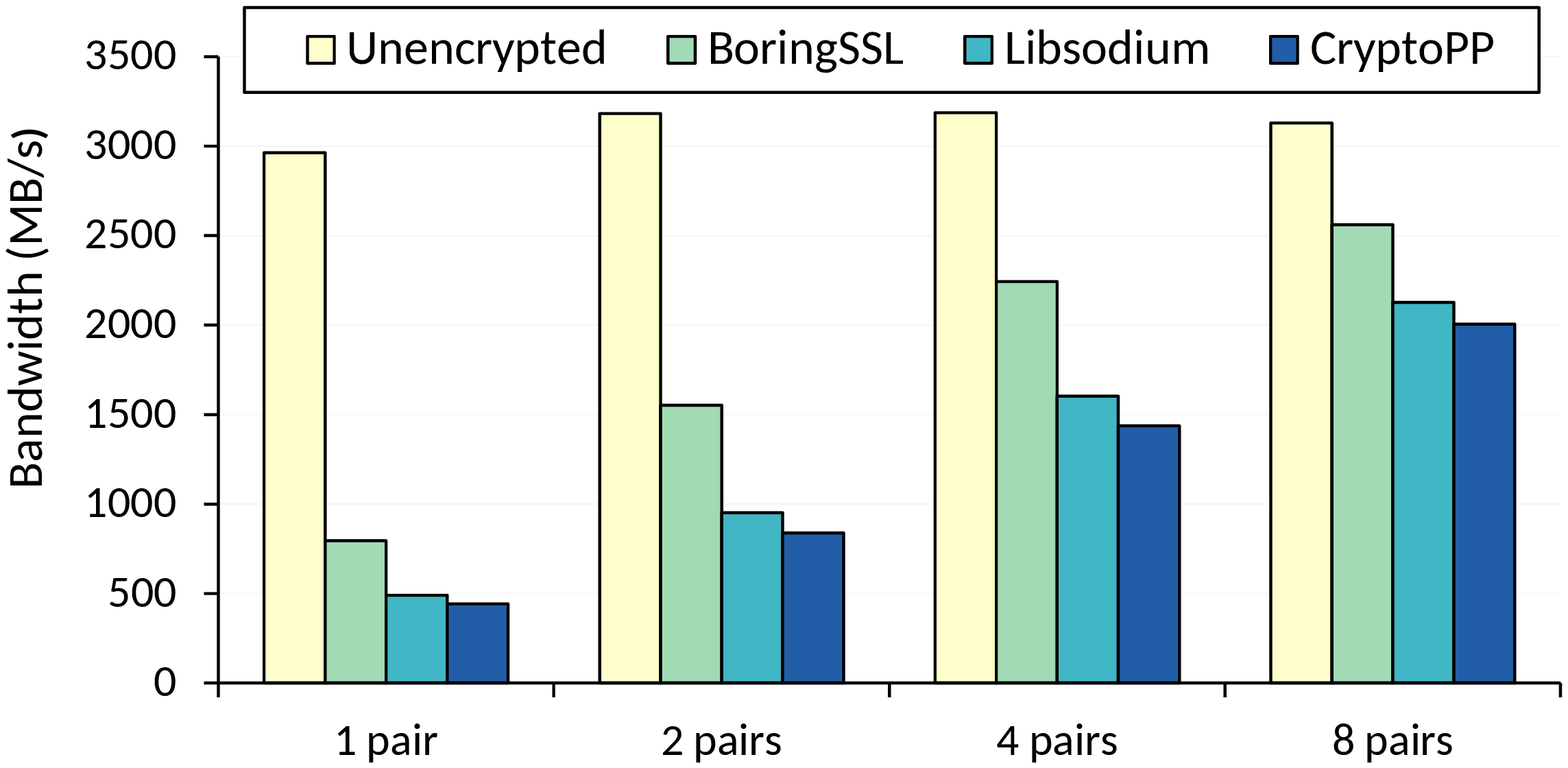}%
}
\\
\subfloat[2MB-messages]{
  \includegraphics[width=0.4\textwidth]{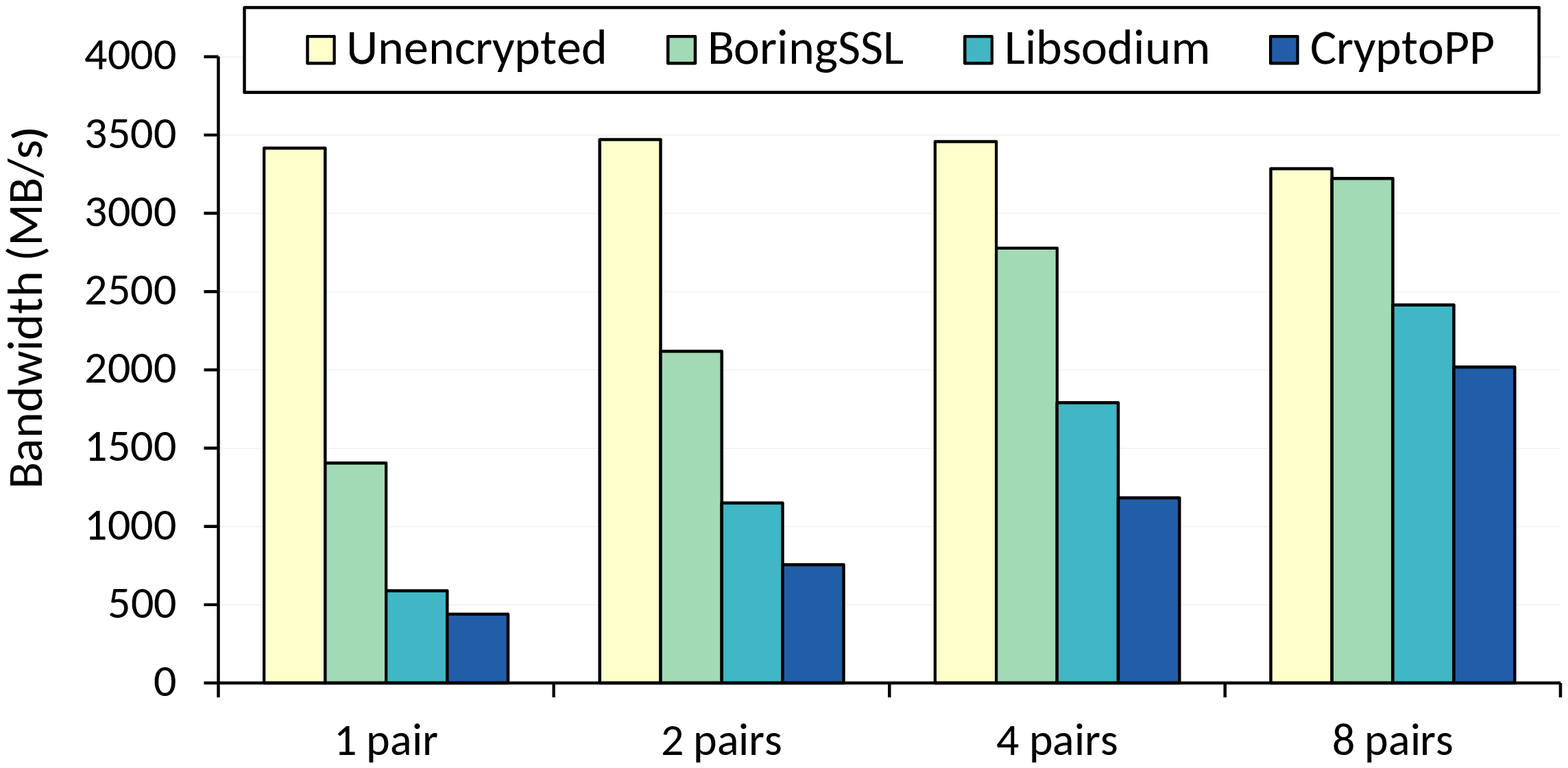}%
}
\captionsetup{singlelinecheck=false}
\caption{OSU Multiple-Pair average throughput on Infiniband. 
}
	\label{fig:multipair_Infiniband_1B}
		\label{fig:multipair_Infiniband_2M}
\vspace{-1.5ex}
\end{figure}

\heading{OSU Multiple-Pair Bandwidth.} 
The Multiple-Pair performance of the baseline and the encrypted MPI libraries, for 1B, 
16KB, and 2MB messages, is shown in Figure~\ref{fig:multipair_Infiniband_1B}. 

Like the Ethernet setting, for medium and large messages, although the encryption overhead is
substantial when there is only one pair of communication, when the number of pairs
increases, the throughput of encrypted MPI libraries is much closer to the baseline
throughput. However, for medium message size (say 16KB), 
even when there are eight communication flows, BoringSSL
only achieves 2561 MB/s, which is just 81.8\% of the baseline throughput of 3128 MB/s. 
This gap is due to the speed difference between Ethernet and Infiniband.

For small messages, the trend is at first similar to that of the Ethernet setting, 
but when the number of pairs increases from 4 pairs to 8 pairs, 
the baseline throughput is throttled, probably due to network contention. 
This decrease also happens for medium and large messages, 
but not as conspicuously as the case of small messages. 
Due to the plummeting of the baseline throughput, for 8 pairs and one-byte messages, 
BoringSSL's overhead is just 29.91\%, 
whereas for 4 pairs, its overhead is 308.33\%.

\begin{figure}[t]
\centering
\subfloat[$\EncBcast$]{
  \includegraphics[width=0.4\textwidth]{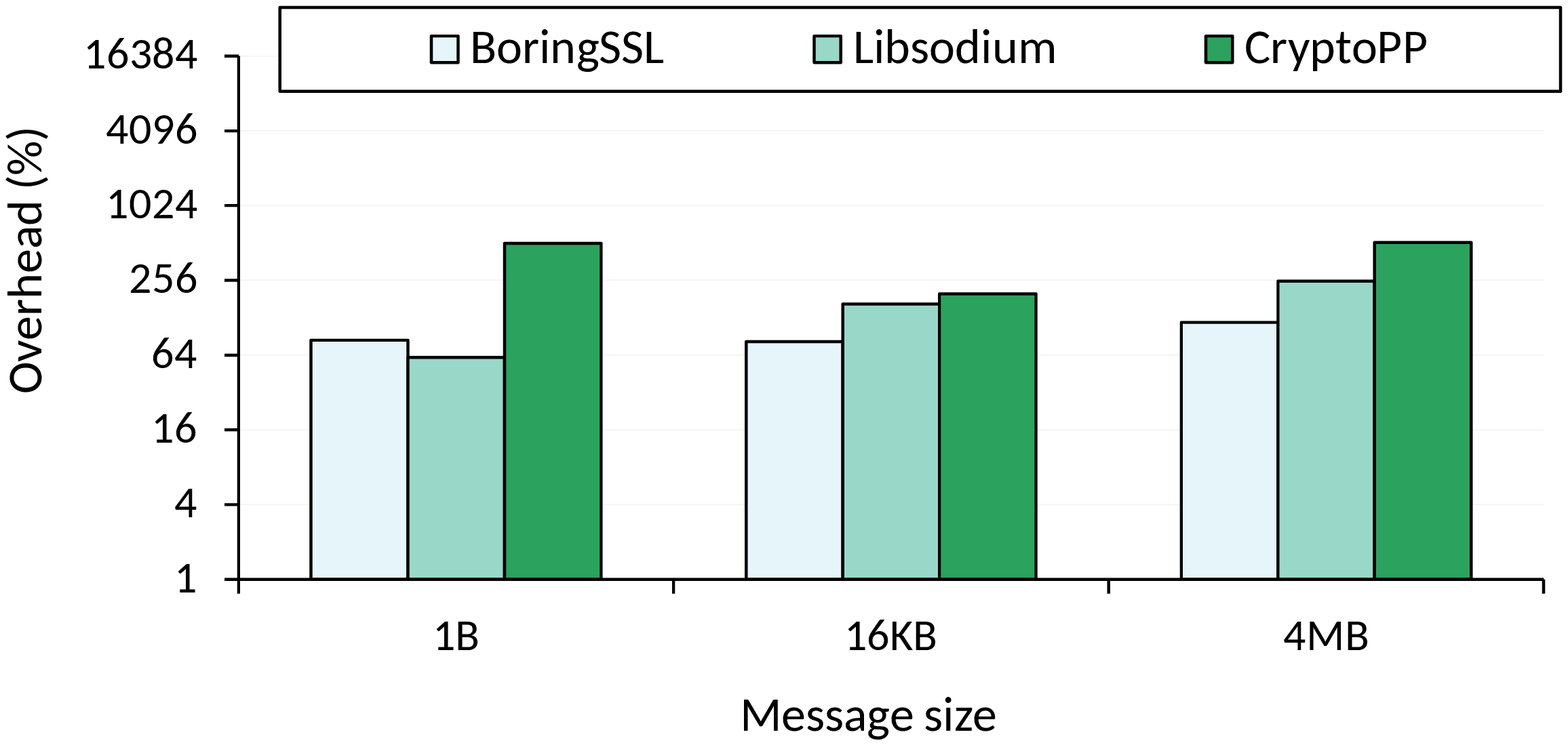}%
}
\\
\subfloat[$\EncAlltoall$]{
  \includegraphics[width=0.4\textwidth]{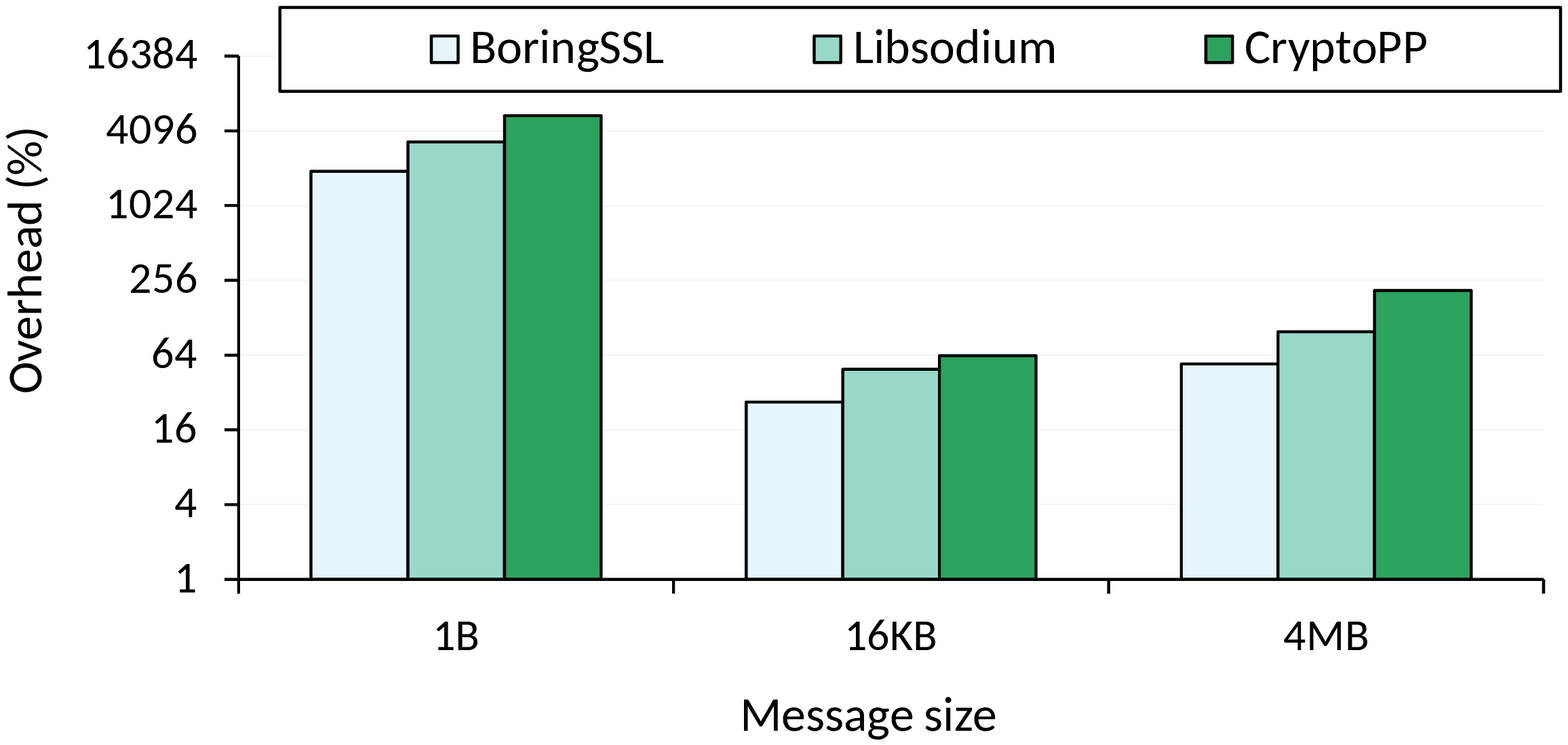}%
}
\caption{Encryption overhead (256-bit key), drawn in log scale, for 
$\EncBcast$ (left) and $\EncAlltoall$(right) on Infiniband. 
}
\label{fig:Bcast_IB}
	\label{fig:Alltoall_IB}
		\vspace{-1.5ex}
\end{figure}

\heading{Collective Communication.} The average running time of $\EncBcast$ 
and $\EncAlltoall$  
for the 64-rank and 8-node setting,  is shown in 
Tables~\ref{tab:bcast_infiniband} and \ref{tab:alltoall_infiniband}
respectively. The trend, as illustrated in \figref{fig:Bcast_IB}, 
is similar to that of the Ethernet setting, 
but the overhead is much worse, because Infiniband latency is lower.

\medskip 

\begin{minipage}{\linewidth}
\centering
\captionsetup{justification=centering, labelsep=newline}
\captionof{table}{Average Timing of $\EncBcast$ ($\mu$s), with 256-bit 
Key on Infiniband.} 
\label{tab:bcast_infiniband} 
\begin{tabular}{ C{.7in}  *2{C{.45in}} C{.8in}}\toprule[1.25pt]	
 &  \textbf{1B} & \textbf{16KB} & \textbf{4MB}   \\ \midrule
\textbf{Unencrypted} &  4.14 & 28.58 & 3,780.27  \\ \midrule
\textbf{BoringSSL} &   7.64 & 52.08 & 8,204.73\\ \midrule
\textbf{Libsodium} &  6.68 & 75.81 & 13,294.35\\ \midrule
\textbf{CryptoPP}  &  25.25 & 85.43 & 23,344.63\\
\bottomrule[1.25pt]
\end{tabular}
\medskip
\end{minipage}

\medskip 

\begin{minipage}{\linewidth}
\centering
\captionsetup{justification=centering, labelsep=newline}
\captionof{table}{Average Timing of $\EncAlltoall$ ($\mu$s),  with 256-bit 
Key on Infiniband.} 
\label{tab:alltoall_infiniband} 
\begin{tabular}{ C{0.9in}  *2{C{.45in}} C{.8in}}\toprule[1.25pt]
 &  \textbf{1B} & \textbf{16KB} & \textbf{4MB}   \\ \midrule
\textbf{Unencrypted} &  21.48 & 5,352.84 & 657,145.51  \\ \midrule
\textbf{BoringSSL} & 435.70 & 6,789.17   & 1,013,896.50 \\ \midrule
\textbf{Libsodium} &  736.29 & 7,977.41 & 1,305,389.60\\ \midrule
\textbf{CryptoPP}  &  1,187.75 & 8,744.08 & 2,049,864.38 \\
\bottomrule[1.25pt]
\end{tabular}
\medskip
\end{minipage}

\smallskip

\heading{NAS benchmarks.} The results of NAS benchmarks for Infiniband are shown in Table~\ref{tab:NAS_IB}. 
Here the overheads of BoringSSL, Libsodium, and CryptoPP are 17.93\%, 24.27\% and 29.41\% respectively. 
CryptoPP's overhead in Infiniband is slightly less than that in Ethernet, because the compiler in the former
setting uses more aggressive optimizations than its Ethernet counterpart, which drastically improves 
the performance of CryptoPP, as shown in Figures~\ref{fig:enc_dec_gcc_throughput} and~\ref{fig:enc_dec_mvapich}. 
These results again reiterate our thesis that even in very fast networks, encryption overheads may not be prohibitive
for practical scenarios.

\bigskip 

\begin{figure*}
\begin{minipage}{\linewidth}
\centering
\captionsetup{justification=centering, labelsep=newline}
\captionof{table}{Average Running Time (Seconds) of NAS Parallel Benchmarks, Class C, 64-rank \\ and 8-node, on Infiniband.}
\label{tab:NAS_IB}
\begin{tabular}{p{0.19\linewidth}*7{p{0.067\linewidth}}}
\toprule[1.25pt]
 & \textbf{CG} & \textbf{FT} & \textbf{MG} & \textbf{LU} & \textbf{BT} & \textbf{SP} & \textbf{IS}  \\ \midrule
\textbf{Unencrypted} & 6.55 &  10.00 & 3.59 & 18.36 & 24.56 & 24.20 & 3.04 \\ \midrule
\textbf{BoringSSL} & 8.36 & 10.77 & 4.20 & 19.73 & 33.35 & 26.87 & 3.20 \\ \midrule
\textbf{Libsodium} &  9.87 & 11.52 & 4.28 & 20.04 & 34.62 & 28.55 & 3.33 \\ \midrule
\textbf{CryptoPP} & 10.47 & 11.89 & 4.41 & 22.82  & 34.96 & 28.97 & 3.35 \\
\bottomrule[1.25pt]
\end{tabular}
\end{minipage}
\end{figure*}

\section{Modeling Encrypted MPI Point-to-Point Communication}

To further understand the performance of encrypted MPI communications, we develop
performance models that can predict the performance of 
encrypted MPI point-to-point communication
in two settings: single-pair point-to-point communication like
the Ping-Pong test and (concurrent) multiple-pair
point-to-point communication like the OSU Multiple-Pair
benchmark \cite{OSUBM}. Our models extend the 
Hockney model~\cite{Hockney94} and the max-rate model~\cite{gropp2016modeling}, and are 
able to predict the performance of 
encrypted MPI point-to-point communication in the two settings accurately. 
We will describe the models and validate them with measured data in \secref{sec:perf}, 
on both Ethernet and Infiniband. 
We show that the parameters 
for our models can be obtained by independent benchmarking of the MPI and cryptographic libraries.
Using the models, we reason about the performance of encrypted MPI point-to-point 
communication in different system settings and suggest potential optimization techniques
to improve the performance of encrypted MPI communication.

\subsection{Modeling Single-pair Encrypted MPI Point-to-point Communication}

\noskipheading{Modeling communication}. If encryption and decryption are extremely fast then 
the performance of encrypted communication will converge to that of the conventional MPI communication.
Therefore, let's  begin by recalling how to model unencrypted point-to-point communication. 
There are several models in the literature, such as LoP~\cite{Culler93}
or LoGP~\cite{Alexandrov95}. Among them, a popular one is 
the Hockney model~\cite{Hockney94}, where 
the  time $\Tcomm(m)$ to send or receive a message of $m$ bytes is modeled as
\[ 
\Tcomm(m) = \acomm + \bcomm \cdot m \enspace. 
\] 
Here, $\acomm$ is the network latency, and $\bcomm$ is the inverse of the asymptotic bandwidth. 
We will use different values of the parameters for the eager phase (i.e., message size is smaller than $128$ KB on Ethernet)
and the rendezvous phase (i.e., message size is bigger than $128$ KB on Ethernet), 
since MPI implementations will use different protocols for these phases.

To obtain the parameters of the Hockney model, 
we use  linear least squares  on latency measurements from the Ping-Pong benchmarks in \secref{sec:perf}
to find the best fitting $(\acomm, \bcomm)$.\footnote{
If the linear regression results in a meaningless negative $\acomm$
then we will instead let $\acomm$ be the average latency of Ping-Pong benchmarks for one-byte data, 
and then find the best fitting $\bcomm$ via linear least squares. 
}
The values of those parameters are given in Table~\ref{tab:hockney_param}.

\begin{figure}
\begin{minipage}{\linewidth}
\centering
\captionsetup{justification=centering, labelsep=newline}
\captionof{table}{The Values of Parameters $\acomm$  and $\bcomm$ for Unencrypted MPI One-to-One Communication on 10Gbps Ethernet
  and 40Gbps InfiniBand.}
\label{tab:hockney_param} 
\begin{tabular}{ C{0.6in}  l C{0.65in} C{0.8in}}\toprule[1.25pt]
 &  & {$\acomm$ ($\mu$s) } & { $\bcomm$ ($\mu$s/B)} \\ \hline
 \multirow{2}{*}{\textbf{Ethernet}} & Eager & 32.74 &  $23.7 \times 10^{-4}$ \\ 
  & Rendezvous & 117.30 & $8.63 \times 10^{-4}$  \\ \hline
 \multirow{2}{*}{\textbf{Infiniband}} & Eager & 3.40 & $3.83 \times 10^{-4}$ \\ 
  & Rendezvous & 7.17 & $3.12 \times 10^{-4}$ \\ 
 \bottomrule[1.25pt]
\end{tabular}
\end{minipage}
\end{figure}

\heading{Modeling encryption-decryption.} It is also instructive to study encryption-decryption alone, 
as this is a special case of encrypted communication on a very fast network. 
We model the time $\Ted(m)$ of encrypting and then decrypting a message
of $m$ bytes as
\[
\Ted(m) = \aed + \bed \cdot m \enspace, 
\] 
where $\aed$ is the initial overhead, and $\bed$ is the encryption-decryption rate. To obtain the parameters $\aed$ and $\bed$, 
we again use linear least squares on the latency measurements of the encryption-decryption benchmarks in \secref{sec:perf}. 
The values of those parameters are given in  Table~\ref{tab:ed_param}.

\begin{figure}  
\begin{minipage}{\linewidth}
\centering
\captionsetup{justification=centering, labelsep=newline}
\captionof{table}{The Values of Parameters $\aed$  and $\bed$ for Encryption-Decryption.
Since the Performance of BoringSSL or Libsodium are Essentially the Same Regardless of Whether We Compile Under MPICH-3.2.1 or MVAPICH2-2.3, 
We Only Present Their Parameters Under the Former Compiler. 
}
\label{tab:ed_param} 
\begin{tabular}{ l  C{0.65in} C{0.85in}}\toprule[1.25pt]
  & {$\aed$ ($\mu$s) } & { $\bed$ ($\mu$s/B)} \\ \hline
\textbf{ BoringSSL }  & 0.53 &  $6.90 \times 10^{-4}$ \\ \hline
\textbf{ Libsodium  } & 0.48 & $16.3 \times 10^{-4}$ \\ \hline

 \textbf{ CryptoPP ({\small MPICH}) }  & 5.51 & $34.8 \times 10^{-4}$ \\ \hline
 \textbf{ CryptoPP ({\small MVAPICH}) }  & 5.16 & $ 21.4\times 10^{-4}$ \\ 
 \bottomrule[1.25pt]
\end{tabular}
\end{minipage}
\end{figure}

\heading{Modeling encrypted communication.} An encrypted single-flow 
point-to-point  communication consists of the following sequential operations: 
(i) encrypting the message at the sender, (ii) transmitting the encrypted message from the sender to the receiver, and (iii) decrypting
the ciphertext at the receiver. If we ignore the 28-byte expansion in encryption
then the time $\Tall(m)$ in completing the encrypted communication of an $m$-byte message can be modeled as
\setlength{\arraycolsep}{0.0em}
\begin{eqnarray}
  \Tall(m)&{} \approx {}&\Tcomm(m) + \Ted(m) = (\acomm + \aed)\nonumber\\ 
  &&{+}\:(\bcomm + \bed) \cdot m \enspace.\nonumber 
\end{eqnarray}
\setlength{\arraycolsep}{5pt}

If we let $\aall = \acomm + \aed$ and $\ball = \bcomm + \bed$ then we obtain 
\begin{equation}
\Tall(m) = \aall + \ball \cdot m \enspace. 
\label{eq:hockney_all}
\end{equation}
Intuitively, encrypted communication can be viewed as conventional MPI communication over a private and authenticated channel, 
thus \eqref{eq:hockney_all} is the Hockney model for communication over this network. 
Since this Hockney model takes both communication and encryption into account, we will
call it the enhanced Hockney model.

\smallskip
An advantage of the enhanced Hockney model above is that the parameters $\aall$ and~$\ball$ 
can be obtained without implementing encrypted MPI. 
In fact, one can use available benchmarks of conventional MPI implementations and cryptographic libraries 
to obtain $\acomm, \bcomm, \aed, \bed$, and  compute $\aall \gets \acomm + \aed$
and $\ball \gets \bcomm + \bed$. In other words, the independent benchmarking results 
summarized in Table~\ref{tab:hockney_param} (for MPI libraries) and 
Table~\ref{tab:ed_param} (for cryptographic libraries), 
can directly derive the model parameters 
for the enhanced Hockney model. Consider, for example, 
small messages (Eager protocol) with BoringSSL on InfiniBand.
From Table~\ref{tab:hockney_param}, we have 
$\acomm = 3.40$ and $\bcomm=3.83\times 10^{-4}$;  
from Table~\ref{tab:ed_param}, we have 
$\aed = 0.53$ and 
$\bcomm=6.90\times 10^{-4}$. 
Hence, for the encrypted single-flow
MPI point-to-point communication, the enhanced Hockney model has the parameters
$\aall = \acomm + \aed = 3.40+0.53 = 3.93$ and 
$\ball = \bcomm + \bed = 3.83+6.90 = 10.73$ for small messages. 
The model parameters 
for other combinations of libraries and systems can be derived similarly.

\begin{figure*}
\centering
\subfloat[BoringSSL on Ethernet]{
  \includegraphics[width=0.45\textwidth]{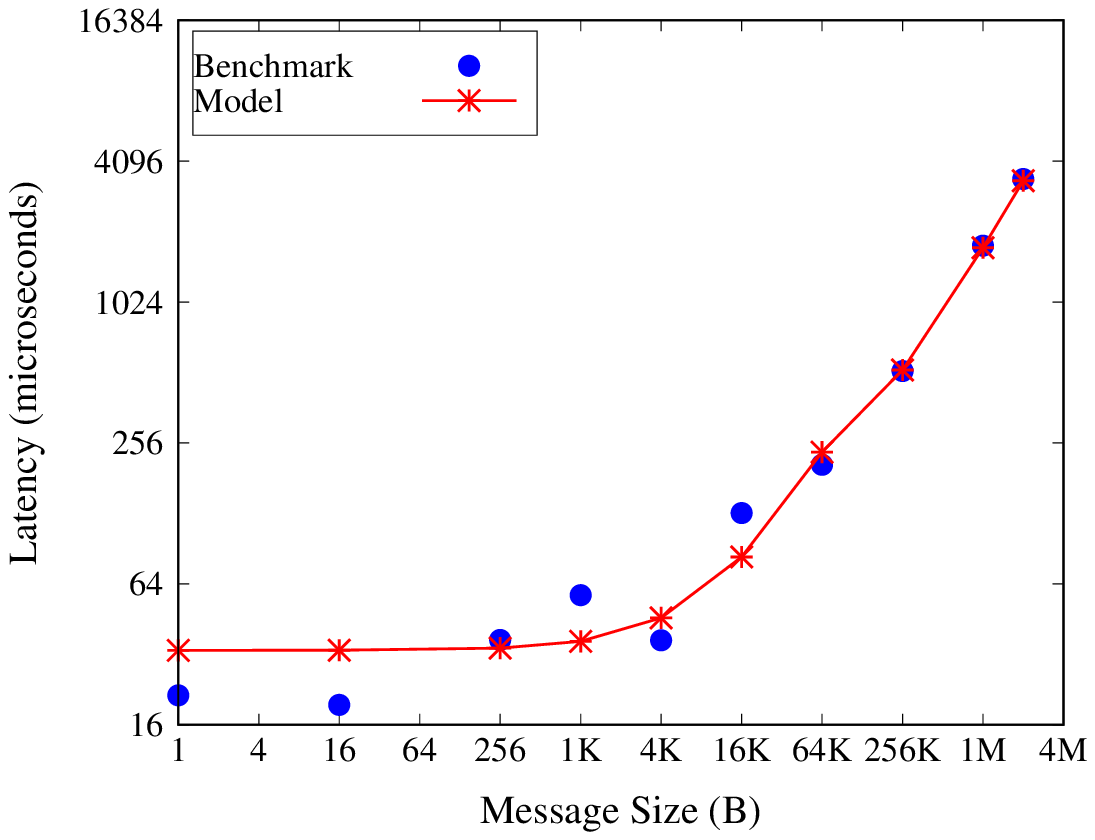}
}
\subfloat[BoringSSL on Infiniband]{
  \includegraphics[width=0.45\textwidth]{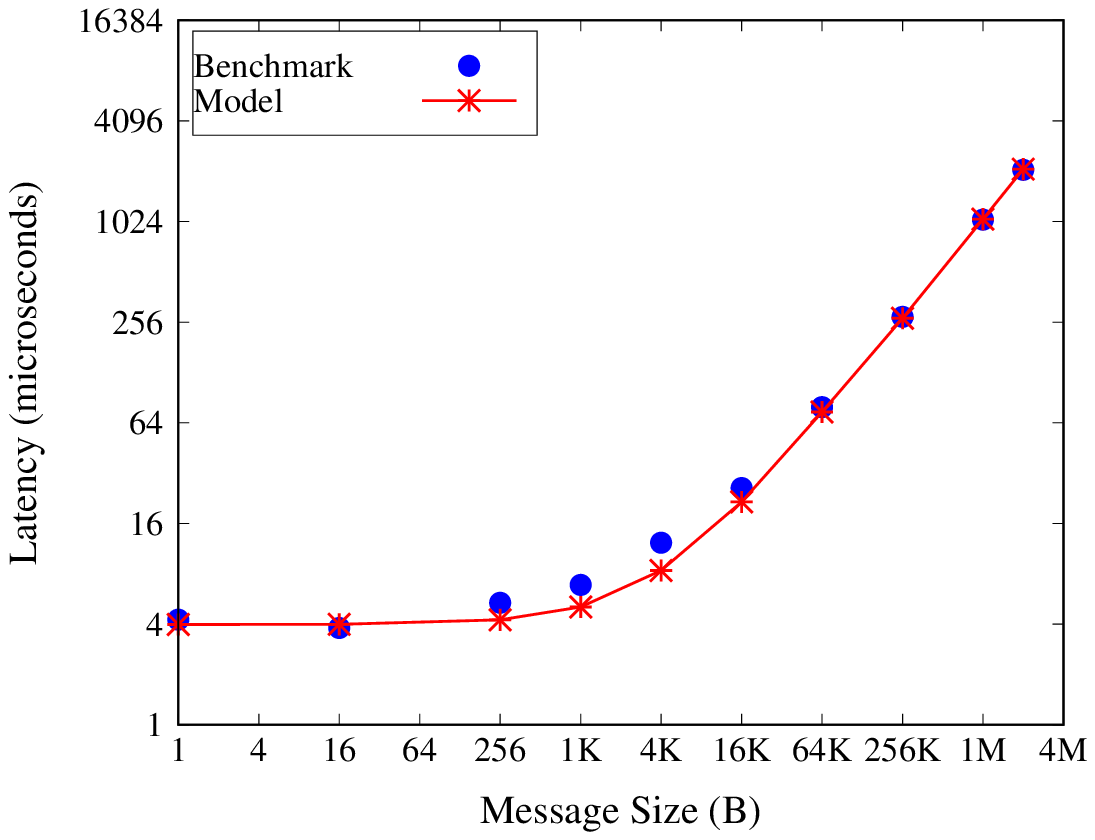}
}
\newline
\subfloat[Libsodium on Ethernet]{
  \includegraphics[width=0.45\textwidth]{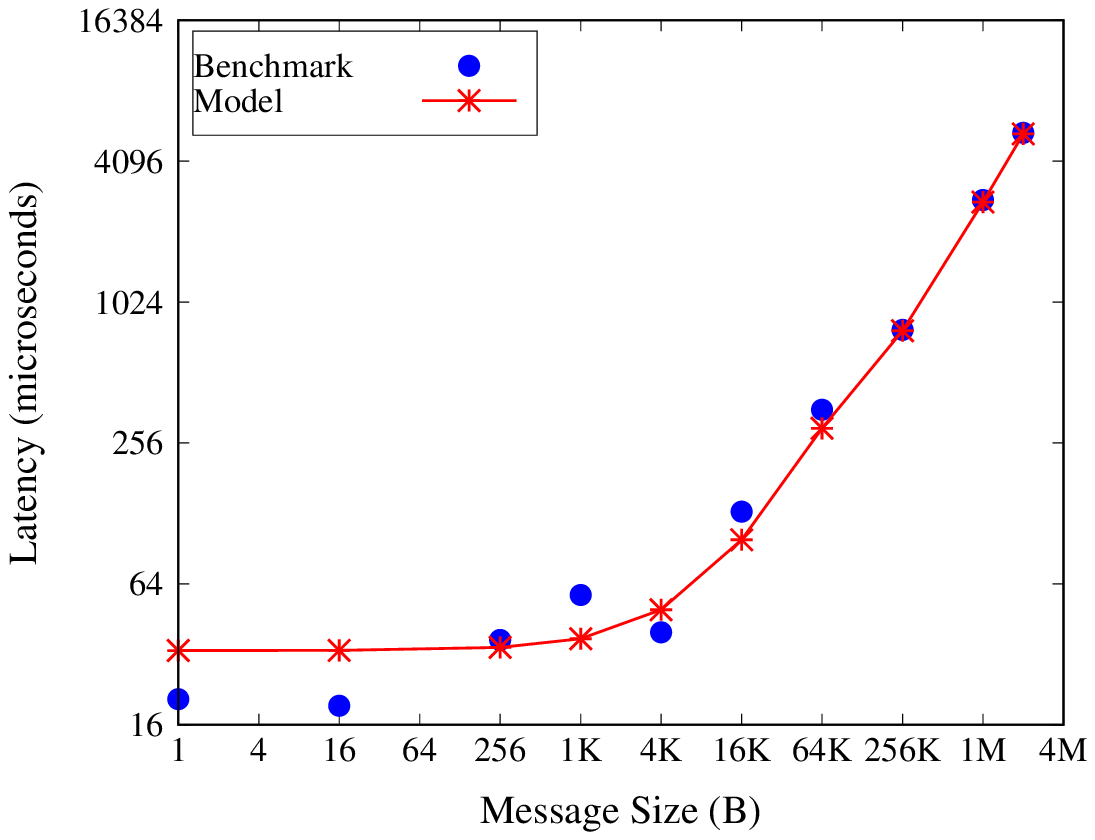}
}
\subfloat[Libsodium on Infiniband]{
  \includegraphics[width=0.45\textwidth]{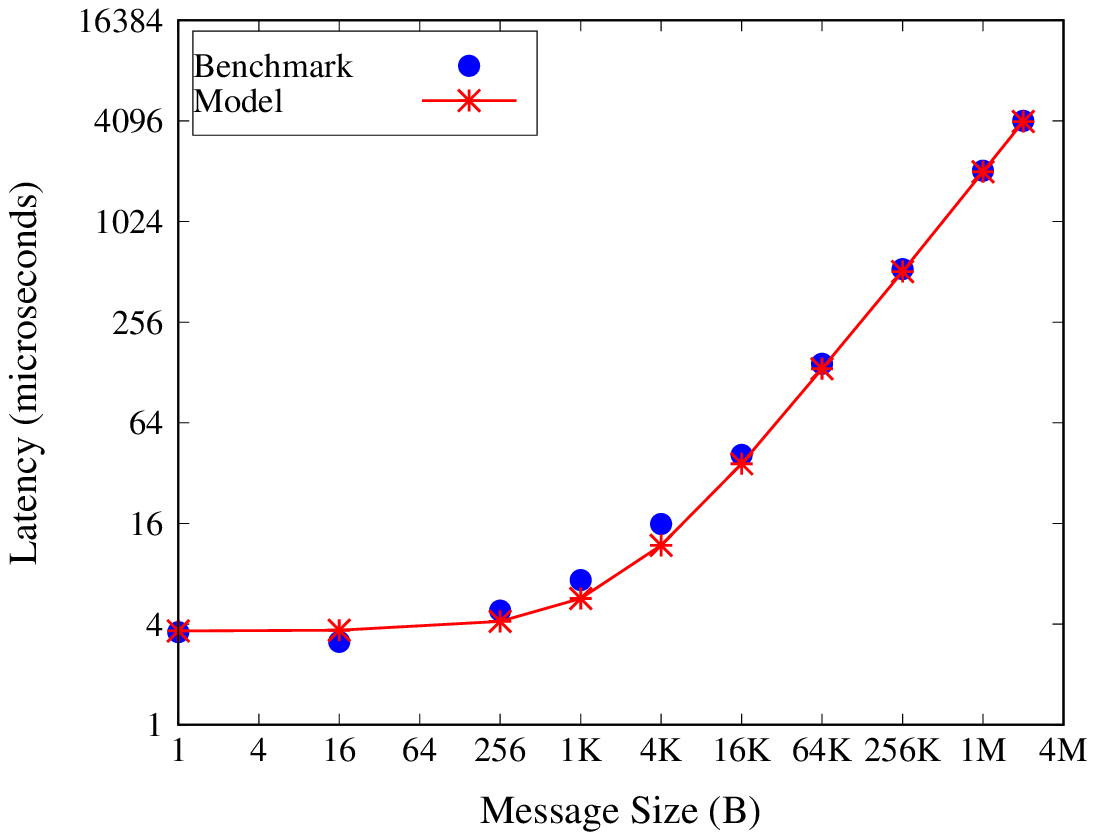}
}
\newline
\subfloat[CryptoPP on Ethernet]{
  \includegraphics[width=0.45\textwidth]{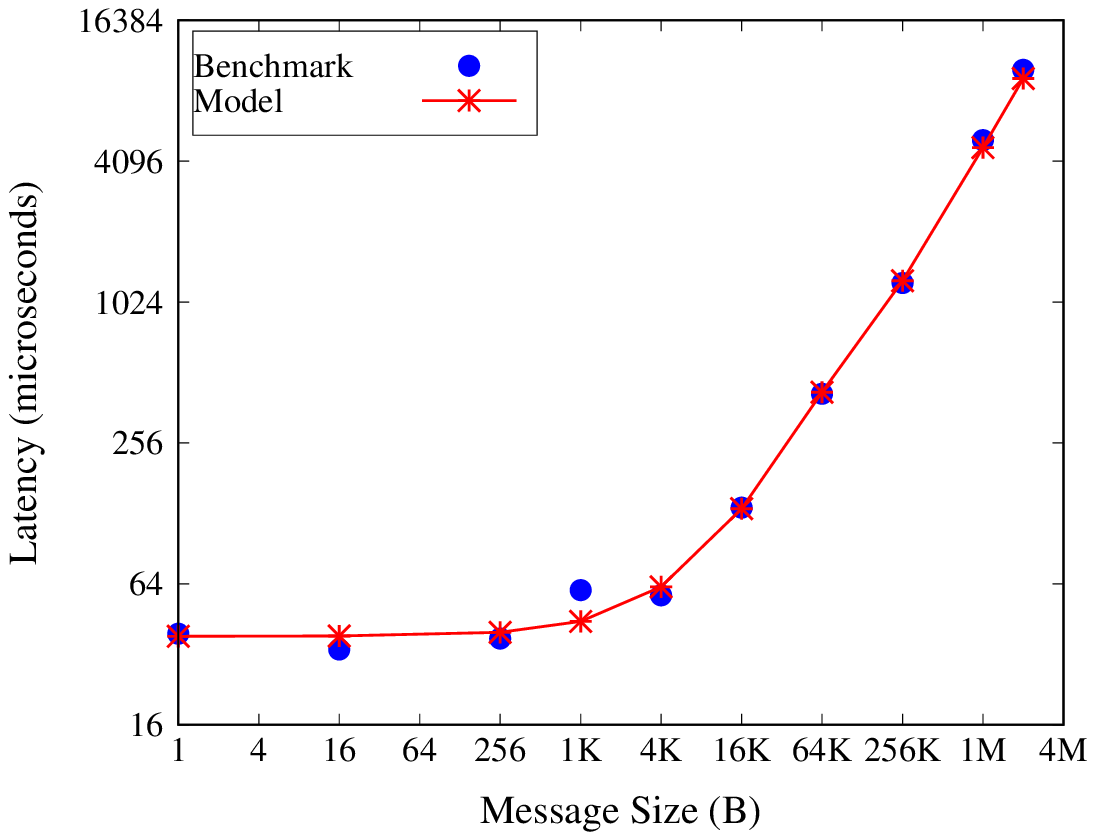}
}
\subfloat[CryptoPP on Infiniband]{
  \includegraphics[width=0.45\textwidth]{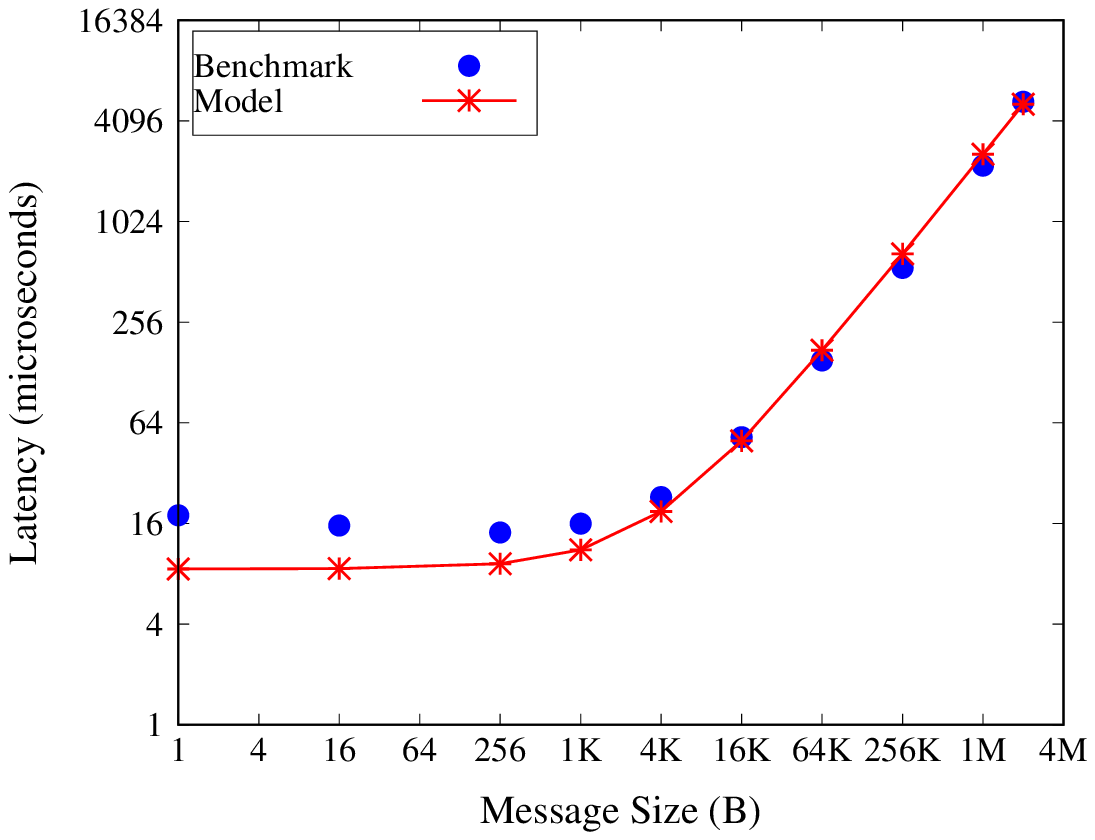}
}
\caption{Latency of encrypted MPI one-to-one communication: benchmark versus model prediction, 
for 256-bit encryption. 
Pictures are drawn in log scale.}
\label{fig:naive_model}
\end{figure*}

Using parameters derived from independent benchmarking of MPI and encryption
libraries, we validate the enhanced Hockney models for single-pair encrypted
MPI communications by comparing the model's predicted performance to the measured
performance of our encrypted MPI libraries with the Ping-Pong benchmark
on different systems. Three encryption libraries, BoringSSL, Libsodium, and CryptoPP,
and two networks, InfiniBand and Ethernet are studied. Results for OpenSSL are
similar to those for BoringSSL and are not reported. 
As shown in \figref{fig:naive_model},
for all configurations, our enhanced Hockney 
models capture the trend of encrypted communication with reasonable accuracy.

\subsection{Modeling Multiple-pair Encrypted MPI Point-to-point 
Communication} \label{sec:multipairs}

We consider the performance of
multiple concurrent pairs of encrypted MPI point-to-point 
communications like the OSU multiple-pair benchmark. 

%
%


\noskipheading{Modeling communication}.
Again, we begin by recalling how to model the 
performance of concurrent unencrypted, point-to-point communications. 
Under the Hockney model~\cite{Hockney94}, 
the time $\Tcomm(k, m)$ of $k$ senders from one node to transmit $m$-byte messages
to $k$ corresponding receivers in another node is 
\[ 
\Tcomm(k, m) = \acomm + \bcomm \cdot km \enspace, 
\] 
where $\acomm$ is the network latency, and $\bcomm$ is the inverse of the network bandwidth. 
We use different values of the parameters for the eager phase (meaning message size 
is smaller than 128KB on Ethernet) 
and the rendezvous phase (meaning message size is bigger than 128KB on Ethernet).

We use linear least-squares 
on latency measurements from the (unencrypted) Multiple-Pair benchmarks in \secref{sec:perf} to obtain the best fitting $(\acomm, \bcomm)$.
The values of those parameters are given in Table~\ref{tab:maxrate_param}.
\footnote{Since OSU Multiple-Pair benchmark uses non-blocking sends
and Ping Pong uses blocking sends, 
the parameter $\bcomm$ of the former is smaller than that of the latter.  
Moreover, as OSU Multiple-Pairs benchmark sends 64 messages back-to-back, its latency $\acomm$ is  an order of magnitude smaller  than that of the Ping Pong benchmark.}

\begin{figure}
\begin{minipage}{\linewidth}
\centering
\captionsetup{justification=centering, labelsep=newline}
\captionof{table}{The Values of Parameters $(\acomm, \bcomm)$ for Unencrypted Multiple-Pair Experiments on 10Gbps Ethernet
  and 40Gbps Infiniband.}
\label{tab:maxrate_param} 
\begin{tabular}{ C{0.58in}  l C{0.65in} C{0.85in} } \toprule[1.25pt]
 &  & {$\acomm$ ($\mu$s) } & { $\bcomm$ ($\mu$s/B)}   \\ \hline
 \multirow{2}{*}{\textbf{Ethernet}} & Eager & 3.84 & $8.11 \times 10^{-4}$     \\ 
                                    & Rendezvous & 16.35 & $8 \times 10^{-4}$ \\ \hline
 \multirow{2}{*}{\textbf{Infiniband}} & Eager & 1.02 & $2.88 \times 10^{-4}$ \\ 
                                      & Rendezvous & 2.38 &  $2.78 \times 10^{-4}$ \\ 
 \bottomrule[1.25pt]
\end{tabular}
\end{minipage}
\end{figure}

\heading{Modeling encryption-decryption.} To understand how encryption performance scales with multiple threads, 
we consider the setting in which there are~$k$ threads, 
each encrypting and then decrypting an $m$-byte message. 
We follow the max-rate model of Gropp et al.~\cite{gropp2016modeling} 
for concurrent point-to-point communications by 
viewing multi-threading encryption-decryption as unencrypted Multiple-Pair 
communication over a very fast network,
and encryption as injecting a message to a network card. 
Specifically, 
the latency $\Ted(k, m)$ is modeled~as 
\[ 
\Ted(k, m) = \aed + \frac{km}{ A + B \cdot (k - 1)} \enspace, 
\] 
where $\aed$ is the initial cost, and $A + B \cdot (k - 1)$ models the encryption rate of $k$ threads, 
capturing the fact that adding more threads will slow down the per-thread performance.

To obtain the training data for deriving the parameters, we run the encryption-decryption benchmark in \secref{sec:perf} for BoringSSL
for $k \in \{1, 2, 4, 8\}$ threads. 
To account for  the pipelining effect of AES-NI and  whether a message fits the L1 cache (32 KB), 
we divide the message size into three  corresponding levels---small (up to 256B), 
moderate (larger than 256B but smaller than 32KB), and large (32KB or more)---and use different parameters $(\aed, A, B)$ for each level. 
To obtain the parameters $(\aed, A, B)$, we use Matlab's non-linear least square on the latency measurements
from the  experiment above to find the best fitting choice.\footnote{Specifically, we use Matlab's command \texttt{lsqnonlin}.} 
The values of $(\aed, A, B)$ are given in Table~\ref{tab:enc_max_param}.  

\begin{figure}  
\begin{minipage}{\linewidth}
\centering
\captionsetup{justification=centering, labelsep=newline}
\captionof{table}{The Values of Parameters $(\aed, A, B)$ for Multi-Threading Encryption-Decryption on BoringSSL.}
\label{tab:enc_max_param} 
\begin{tabular}{ l C{0.7in} C{0.65in} C{0.8in}}\toprule[1.25pt]
 \textbf{Phase} &  {$\aed$ ($\mu$s) } & { $A$ (B/$\mu$s)}  & $B$ (B/$\mu$s) \\ \hline
 Small & 1.8&  $888.5$ & $0$ \\ \hline  
 Moderate & 2.66 &  $1764$ & $4135$ \\ \hline  
 Large & 3.44 &  $1502.21$ & $1262.59$ \\ 
 \bottomrule[1.25pt]
\end{tabular}
\end{minipage}
\end{figure}

\heading{Modeling encrypted communication.}
Multiple-pair MPI communication is more complicated than
single-pair ping-pong communication in that there are different ways that the
multiple-pair communication can be performed. Hence the models for
multiple-pair communication will depend on the ways that the multiple-pair
communication is performed. We will use the model for
the OSU multiple-pair benchmark to illustrate how multiple-pair encrypted
MPI point-to-point communication can be obtained from independent benchmarking
results for MPI and encryption libraries, that is, from
$\Tcomm(k, m)$ and $\Ted(k, m)$.

In the OSU multiple-pair benchmark, the performance is measured for iterations
of 64-round loops. For each iteration, each sender encrypts a message of $m$
bytes and then sends the ciphertext to its receiver via a non-blocking send; 
with our encrypted library implementation, each receiver waits for all 64 ciphertexts 
to arrive and decrypt them. In modeling this multiple-pair communication,
we make the following assumptions: 
\begin{itemize}
\item The encryption time and decryption time for a message are the same.\footnote{In modern machines that support AES-NI and CLMUL instruction set, AES-GCM encryption is often slightly faster than decryption, 
but the speed difference is small.
} 

\item The transmission of the $i$-th message (via non-blocking send) and the encryption of the $(i + 1)$-th message can happen concurrently. 
Likewise, the receiver can simultaneously decrypt a ciphertext of the $i$-th message while receiving the ciphertext of the $(i + 1)$-th message. 
\end{itemize}

In this communication, each sender will first encrypt the $m$ bytes data, which takes
$\frac{\Ted(k, m)}{2}$ time (half of the encryption and description time), after that,
there are 63 overlapped encryptions and communications, which takes 
$63 \cdot \max\Bigl\{ \frac{\Ted(k, m)}{2}, \Tcomm(k, m) \Bigr\}$ time, and one last communication,
that takes $\Tcomm(k, m)$ time. Hence, the encryption and sending time can be modeled as 

{
\small
\begin{eqnarray} 
 \!\!\!\!\!\!\!\!\!\!\!\!&&\frac{\Ted(k, m)}{2} + 63 \cdot \max\Bigl\{ \frac{\Ted(k, m)}{2}, \Tcomm(k, m) \Bigr\} + \Tcomm(k, m)\nonumber  \\
 \!\!\!\!\!\!\!\!\!&&\approx 64  \cdot \max\Bigl\{ \frac{\Ted(k, m)}{2}, \Tcomm(k, m) \Bigr\} \enspace.\nonumber 
\end{eqnarray}
}


Since the receiver waits for all 64 ciphertexts to arrive and decrypt them, 
the total time for a 64-round loop in the encrypted Multiple-Pair experiment can be 
modeled as
\[ 
64  \cdot \max\Bigl\{ \frac{\Ted(k, m)}{2}, \Tcomm(k, m) \Bigr\} + 64 \cdot  \frac{\Ted(k, m)}{2}
\] 
and thus the average time $\Tall(k, m)$ for a single round is
\setlength{\arraycolsep}{0.0em} 
\begin{eqnarray}
\max\left\{ \frac{\Ted(k, m)}{2}, \Tcomm(k, m) \right\} + \frac{\Ted(k, m)}{2}  \enspace.\nonumber  
\end{eqnarray}
\setlength{\arraycolsep}{5pt}
Notice that the model for multiple-pair encrypted MPI point-to-point
communication, $\Tall(k, m)$ is derived from $\Tcomm(k, m)$ and
$\Ted(k, m)$. Hence the model parameters for multiple-pair communication
can also be derived from independent benchmarking results for the MPI
library  and encryption libraries.

\begin{figure}
\centering
\subfloat[1 pair]{
  \includegraphics[width=0.4\textwidth]{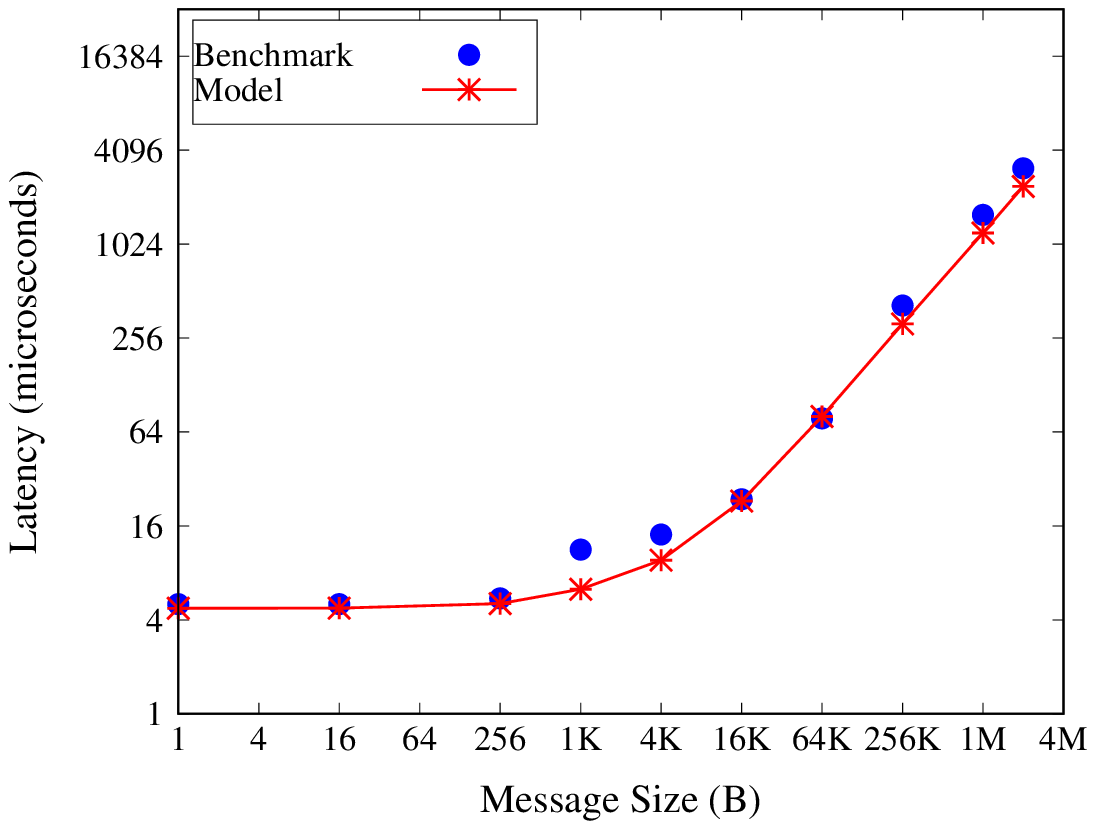}
}
\\
\subfloat[2 pairs]{
  \includegraphics[width=0.4\textwidth]{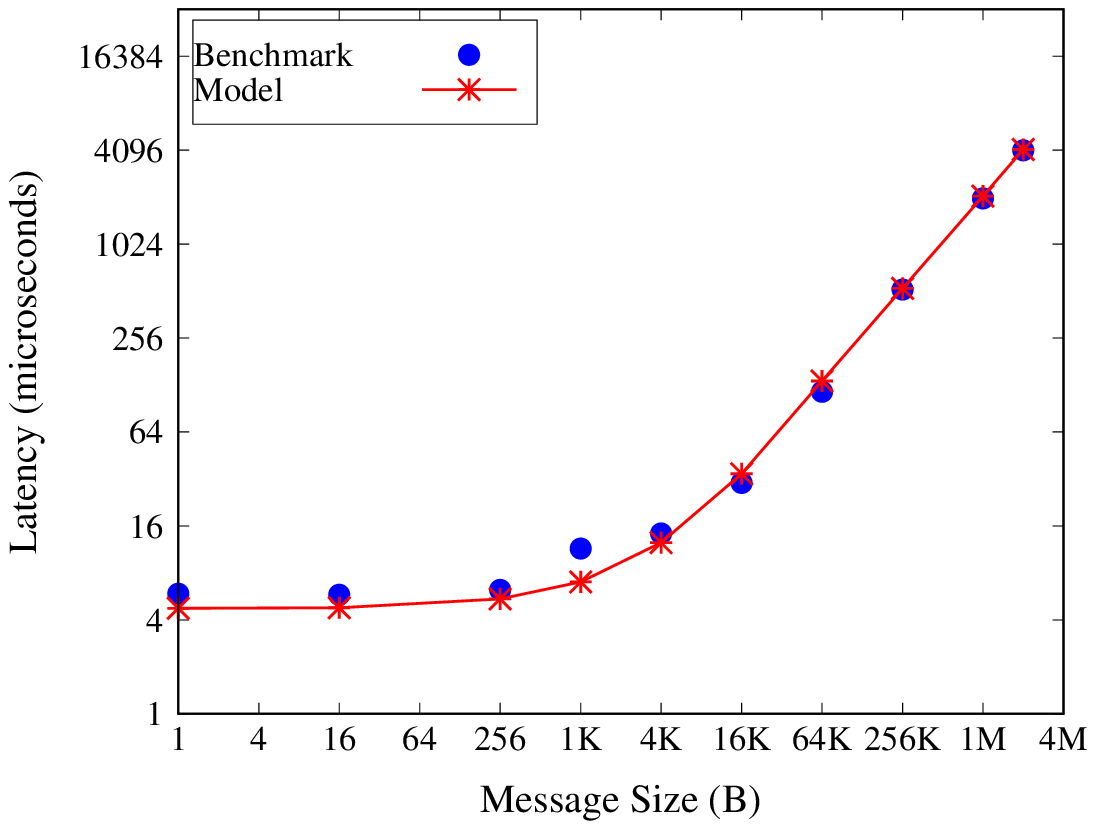}
}
\\
\subfloat[4 pairs]{
  \includegraphics[width=0.4\textwidth]{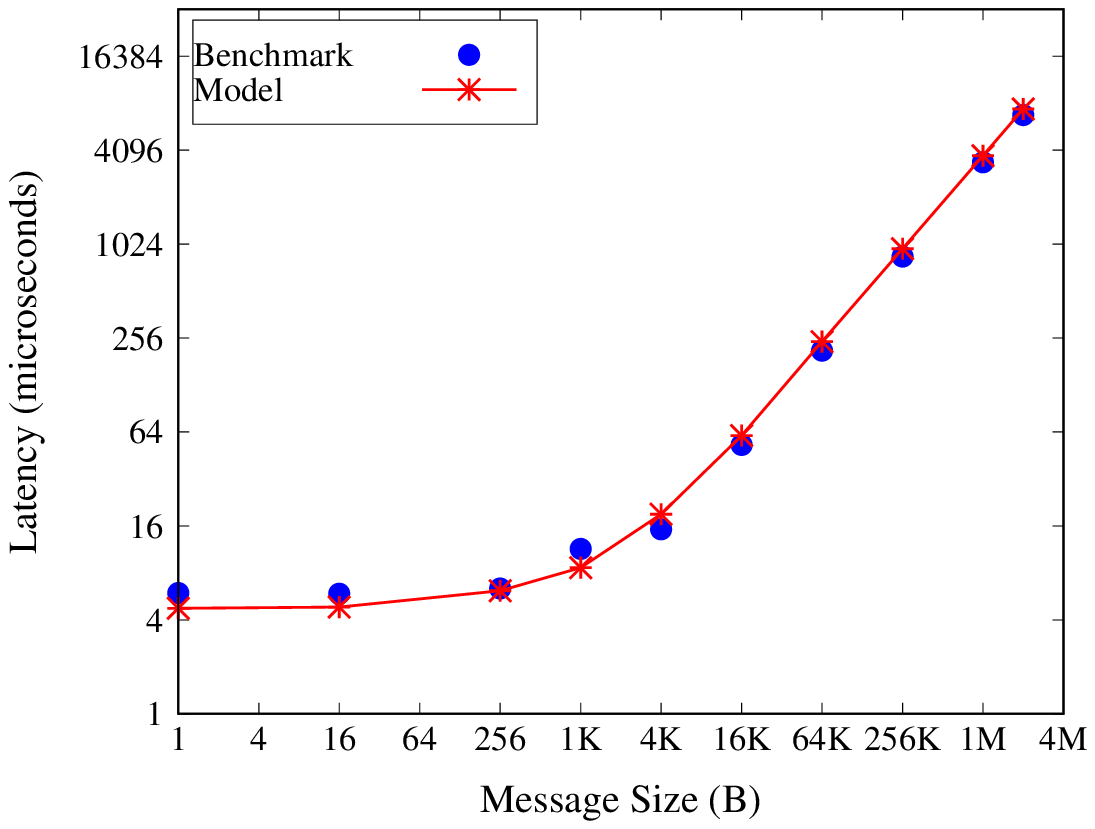}
}
\\
\subfloat[8 pairs]{
  \includegraphics[width=0.4\textwidth]{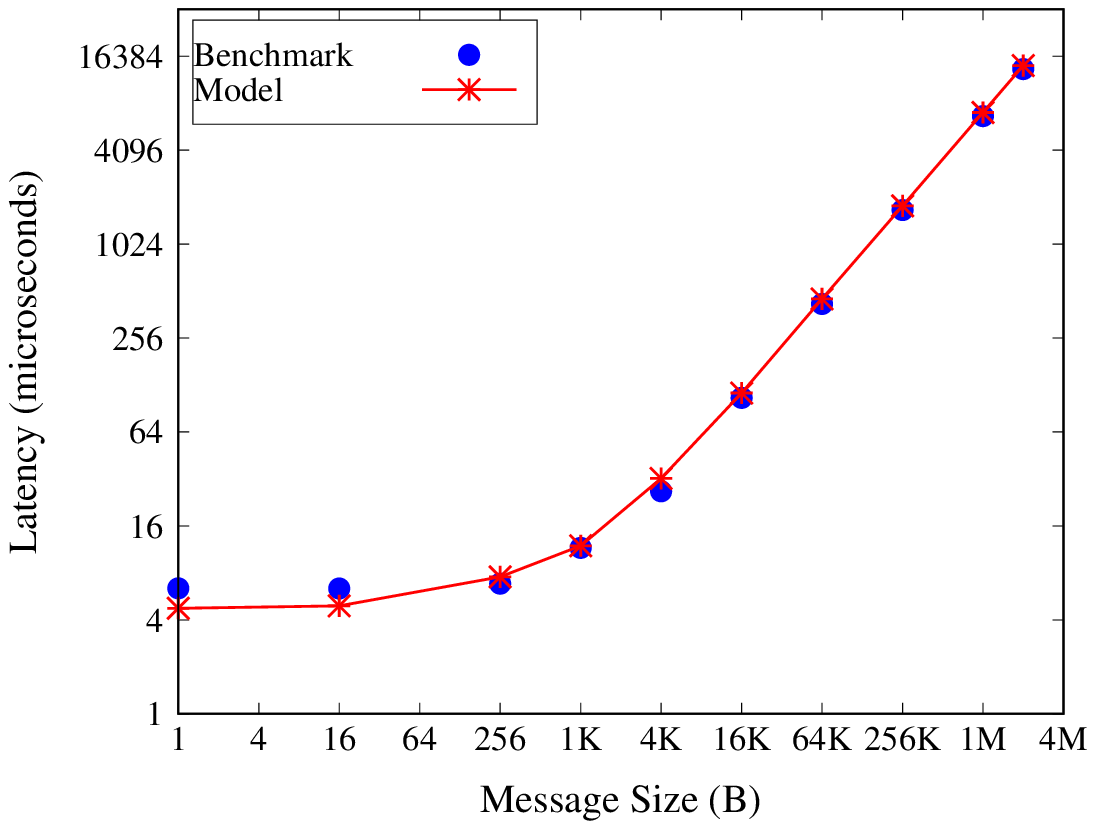}
}
\caption{Latency of encrypted Multiple Pairs on Ethernet: benchmark versus model prediction, 
for 256-bit encryption on BoringSSL. 
Pictures are drawn in log scale.}
\label{fig:multi_Ethernet}
\end{figure}

\begin{figure*}
\centering
\subfloat[1 pair]{
  \includegraphics[width=0.4\textwidth]{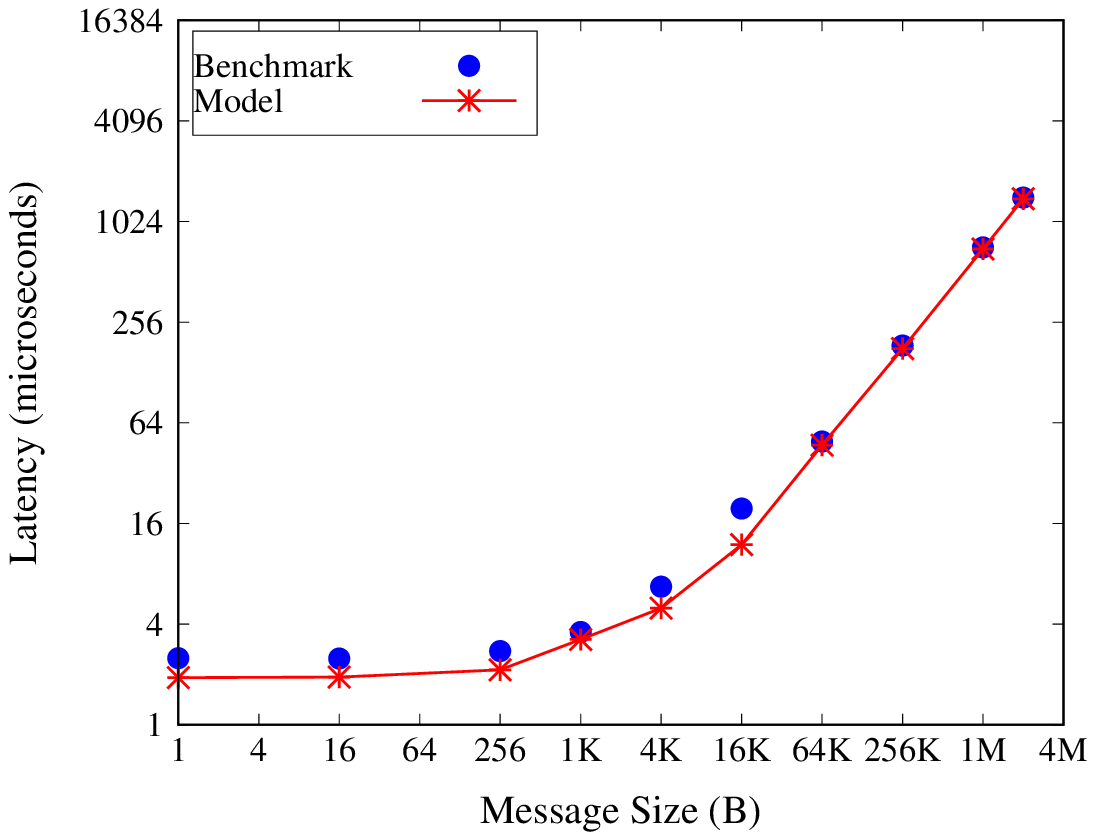}
}
\subfloat[2 pairs]{
  \includegraphics[width=0.4\textwidth]{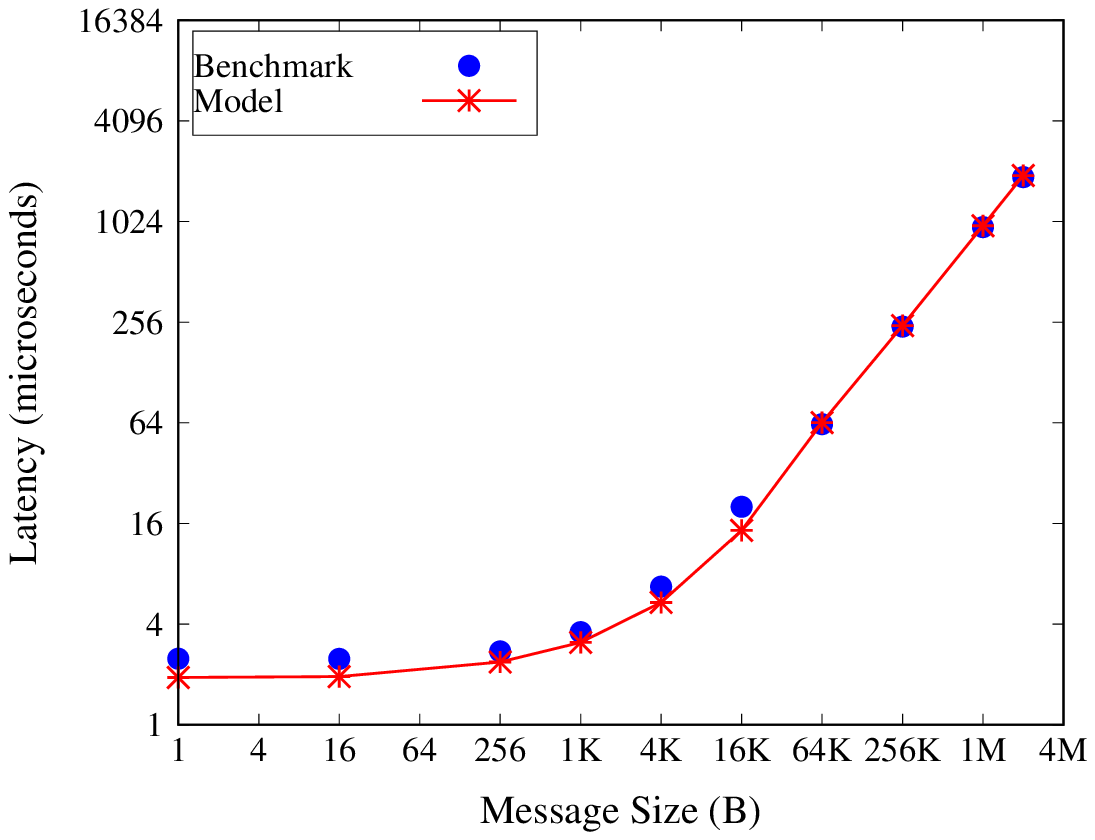}
}
\\
\subfloat[4 pairs]{
  \includegraphics[width=0.4\textwidth]{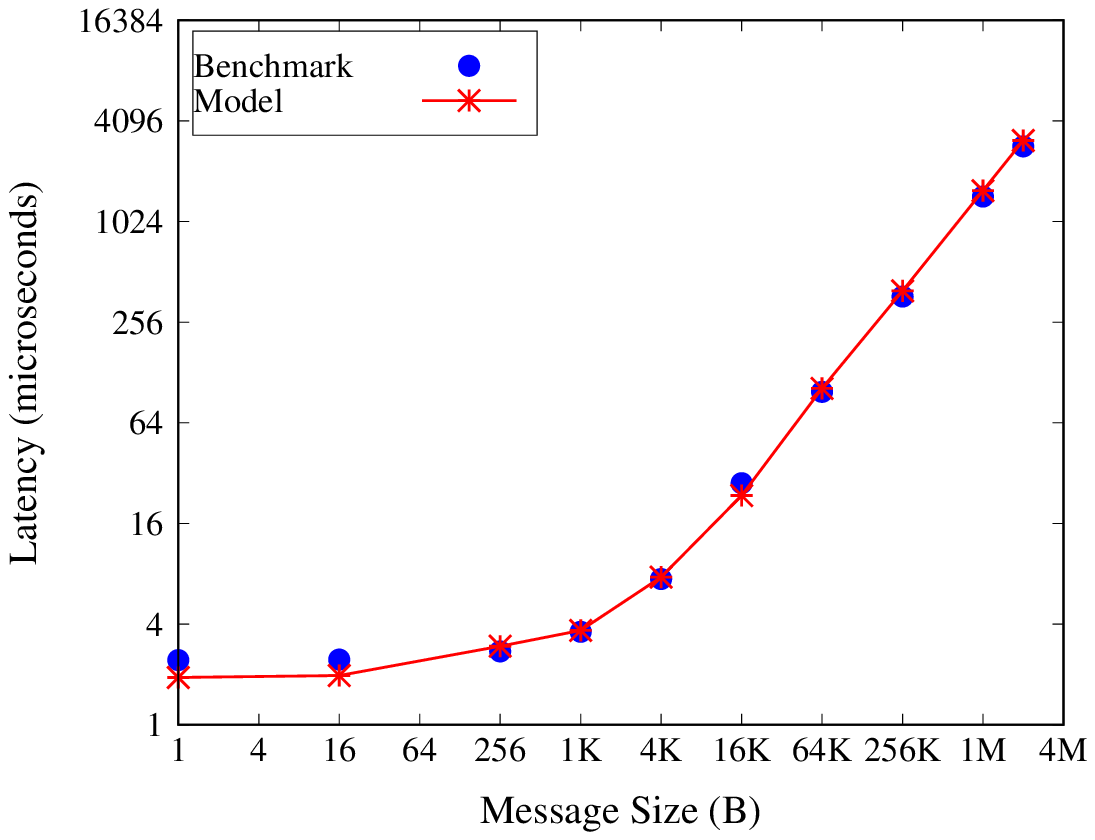}
}
\subfloat[8 pairs]{
  \includegraphics[width=0.4\textwidth]{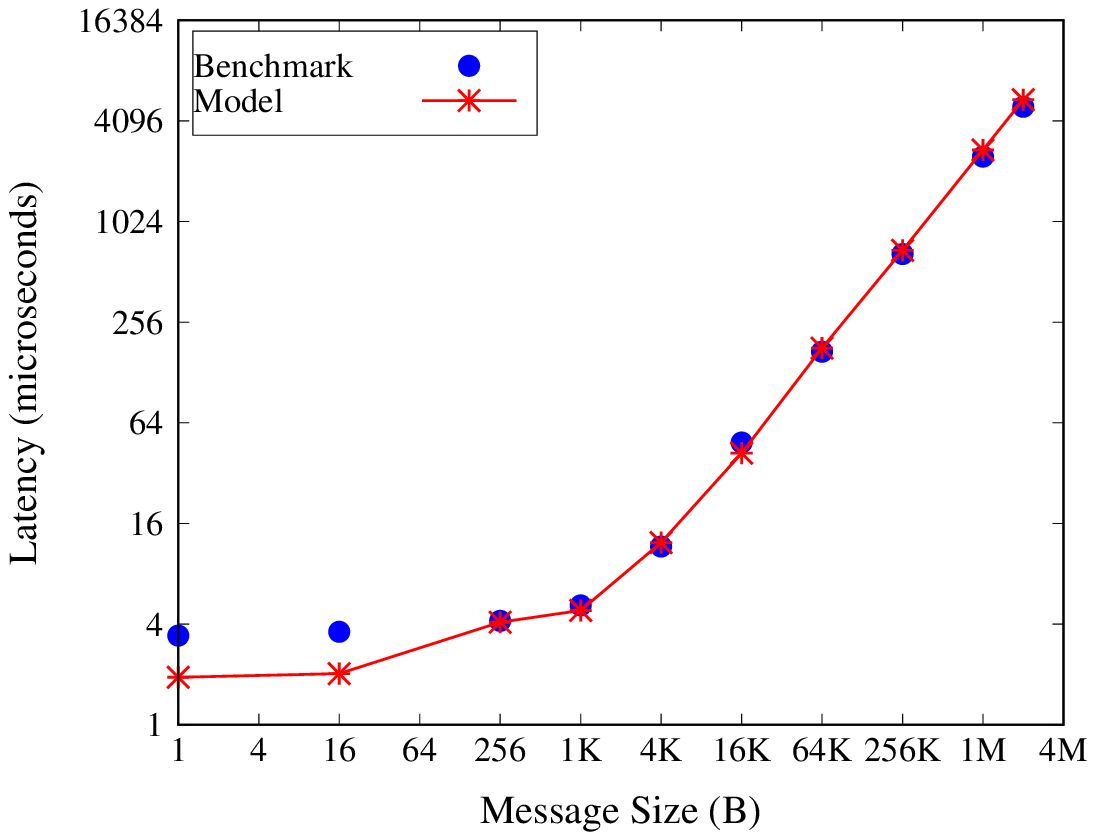}
}
\vspace{-1ex}
\caption{Latency of encrypted Multiple Pairs on Infiniband: benchmark versus model prediction, 
for 256-bit encryption on BoringSSL. 
Pictures are drawn in log scale.}
\label{fig:multi_IB}
\vspace{-1ex}
\end{figure*}

We validate the model by comparing the predicted performance
from the model to the measured
performance using the OSU Multiple-Pair benchmark \cite{OSUBM} with the three 
encryption libraries and two networks (Ethernet and InfiniBand). 
\figref{fig:multi_Ethernet} shows the results for Ethernet
and \figref{fig:multi_IB} shows the results for InfiniBand.

\subsection{Discussion}

An advantage of our models for both single-pair and
multiple-pair communications is that the parameters can be obtained from the 
available benchmarks of conventional MPI and encryption,
meaning that using these models, one does not need to implement
encrypted MPI in order to understand encrypted MPI communication
performance. As shown earlier, these models capture the performance
trend and predict the performance with reasonable accuracy for different
libraries and networks. Hence, the models allow us to reason
about the performance of encrypted MPI communications. 

We will use examples to show that the models give simple and
intuitive explanation of the performance results across
different cryptographic libraries and different networks
in Section~\ref{sec:perf}.
Consider for example the single-pair communication.
For large messages, $\Tcomm(m)$ and $\Tall(m)$ are
dominated by $\bcomm \cdot m$ and $\ball \cdot m$ respectively, 
and thus the encryption overhead  (for throughput)~is 
\[ 
   \frac{\Tall(m)}{\Tcomm(m)} -1 =\frac{\Ted(m)}{\Tcomm(m)} \approx \frac{\bed \cdot m}{\bcomm \cdot m} = \frac{\bed}{\bcomm} \enspace. 
\] 
For example, the overhead of BoringSSL for large messages is predicted
to be 
\[ \frac{\bed}{\bcomm} =\frac{6.9\times 10^{-4}}{8.0\times 10^{-4}}
\approx 86\%\]
for Ethernet, which is consistent with the
Ping-pong benchmark results
in Figure~\ref{fig:pingpong_ethernet} in \secref{sec:perf}---the measured results have an overhead of 78.3\% for 2MB messages.
For InfiniBand, the overhead of BoringSSL for large messages is predicted
to be 
\[ \frac{\bed}{\bcomm} =\frac{6.9\times 10^{-4}}{3.12\times 10^{-4}}
\approx 221\% \enspace, 
\]
where as the measured results of the Ping-pong benchmark
in Figure~\ref{fig:pingpong_infiniband} in \secref{sec:perf} have
an overhead of 215\% for 2MB messages.

Consider now the multiple-pair communication: 
\[ \Tall(k, m) =  \max\left\{ \frac{\Ted(k, m)}{2}, 
\Tcomm(k, m) \right\} + \frac{\Ted(k, m)}{2} \enspace.\]
There are two situations depending on the relative speed of communication 
and encryption. For a slow network such that 
$\Tcomm(k,m) \ge \frac{\Ted(k, m)}{2}$, 
we have $\Tall(k, m) = \Tcomm(k, m) + \frac{\Ted(k, m)}{2}$, 
and the encryption overhead (for throughput) is 

\[ 
  \frac{\Ted(k, m)}{2 \cdot \Tcomm(k, m)} \approx \frac{1}{2\bcomm\bigl( A + (k - 1) \cdot B \bigr)} \enspace, 
\]
meaning that the overhead is inversely proportional to $2k$:
the larger the value of $k$ (pairs of communications),
the smaller the overhead. Both 10Gbps Ethernet and 40Gbps InfiniBand
has $\Tcomm(k,m) \ge \frac{\Ted(k, m)}{2}$ when $k=8$. 
This overhead trend is observed in Figure~\ref{fig:multipair_ethernet_1B} and
Figure~\ref{fig:multipair_Infiniband_1B}.
Plugging in the values of $A$ and $B$ from Table~\ref{tab:enc_max_param}
and $\bcomm$ from Table~\ref{tab:maxrate_param} will give a reasonable
estimation of the overheads for OSU multiple pair benchmarking results. 

For a faster network such that $\Tcomm(k,m) < \frac{\Ted(k, m)}{2}$,
$\Tall(k, m) \approx \Ted(k, m)$, meaning that the communication cost can be completely
hidden. However, the performance will be bounded by the encryption speed,
which can be much slower than the network speed. 

Clearly, there is a gap between what
single-core encryption can provide and what a high-speed network can consume.
The gap is likely to increase since (1)~we reach the limit of Moore's law and 
single-core computing performance will not drastically improve, and (2)
communication technology is still improving as the network bandwidth
continues to increase. As such, efficient encrypted MPI communication
will be even more challenging in the future. 

The models can give guidance on the techniques to achieve higher encrypted
MPI communication performance as well as the potential benefits of such
techniques. Consider for example to improve the single-pair encrypted performance.
One potential technique is to overlap encryption with communication.
For large messages, this can be achieved by chopping the message
into small segments, then encrypting (and decrypting) each segment,
and using non-blocking communication to overlap communication with encryption
and decryption as in streaming encryption~\cite{C:HRRV15}.
Using this technique, $\Tall(m) \approx \max\{\Tcomm(m), \Ted(m)\}$.
For systems with a slow network such as an 10Gbps Ethernet where $\Tcomm(m) \ge \Ted(m)$,
we have $\Tall(m)\approx \Tcomm(m)$: the encryption overhead
can virtually be completely removed. However, for a faster network such as
40Gbps InfiniBand where $\Ted(m) \ge \Tcomm(m)$, $\Tall(m) \approx \Ted(m)$. Consider
for example, encrypted communication for large message with
BoringSSL on InfiniBand, we have $\bcomm = 3.12\times 10^{-4}$
(Table~\ref{tab:hockney_param})
and $\bed = 6.90\times 10^{-4}$
(Table~\ref{tab:ed_param}), hence $\Ted(m) \approx 2.2\Tcomm(m)$. Thus,
even with the encryption and communication overlapping technique, the
encryption overhead will roughly be $\frac{\Tall(m)}{\Tcomm(m)} -1 \approx 120\%$.
This overhead
is further be improved by using multiple threads to perform concurrent encryption
and decryption. This technique with multiple threads will be more attractive
for future systems since they will likely to have more (idle) cores in each node.

\section{Conclusion}

We considered adding encryption to MPI communications 
for providing privacy and integrity. 
Four widely used cryptographic libraries, OpenSSL, BoringSSL, CryptoPP, and
Libsodium, were studied in this paper. 
We found that the encryption overhead differs drastically across libraries, 
and that BoringSSL (and OpenSSL) achieves
the best performance in most settings. 
Moreover, when individual communication is considered, encryption overhead
can be quite large. However, in practical scenarios when multiple communication
flows are carried out concurrently, the overhead is not significant.
In particular, our evaluation with the NAS parallel benchmarks shows that 
using the best cryptographic library BoringSSL, 
our implementation on average only introduces 
12.75\% overhead on Ethernet and 17.93\% overhead  on  Infiniband.

\smallskip 
We then show that the Hockney model accurately model encrypted Ping Pong performance. 
Moreover, the model parameters can be obtained from benchmark results of a conventional MPI library
and a cryptographic library, meaning that one does not even need to implement encrypted MPI. 
In addition, we show how to decompose the performance of encrypted Multiple-Pair benchmark
into the unencrypted Multiple-Pair cost and the multi-threading encryption cost, 
and model these components via the Hockney model and the max-rate model. 
These models allow us to reason about
the performance of encrypted MPI communication across different cryptographic libraries and
platforms, and estimate the potential performance gain for different
optimization techniques. 

\ifCLASSOPTIONcompsoc
  \section*{Acknowledgments}
\else
  \section*{Acknowledgment}
\fi

We thank Sriram Keelveedhi for helpful discussions
and Prof. Weikuan Yu at Florida State University for providing the computing
infrastructure for software development and performance measurement.
This material is based upon work supported by the National Science Foundation
under Grants CICI-1738912, CRI-1822737, CNS-1453020, and CRII-1755539.
Any opinions, findings, and conclusions or
recommendations expressed in this material are those of the author(s) and do not
necessarily reflect the views of the National Science Foundation.
This work used the Extreme Science and Engineering Discovery Environment
(XSEDE), which is supported by National Science Foundation grant number
ACI-1548562. This work used the XSEDE Bridges resource at the Pittsburgh
Supercomputing Center (PSC) through allocation TG-ECS190004.

\ifCLASSOPTIONcaptionsoff
  \newpage
\fi



%


\bibliography{tdsc}
\bibliographystyle{IEEEtran}

%





\vspace{-0.8in}
\end{document}